\documentclass{aa}
\usepackage{natbib}
\bibpunct{(}{)}{;}{a}{}{,}
\usepackage{txfonts}

\newcommand{\msunyr}{\ensuremath{\mathit{M}_{\odot}{\rm yr}^{-1}}}   
\newcommand{\kms}{\ensuremath{{\rm km\,s^{-1}}}}                   
\newcommand{\msun}{\ensuremath{\mathit{M}_{\odot}}}   
\newcommand{\mini}{\ensuremath{M_{\rm ini}}}                         
\newcommand{\lsun}{\ensuremath{\mathit{L}_{\odot}}}                  
\newcommand{\rsun}{\ensuremath{\mathit{R}_{\odot}}}                  

\newcommand{\lstar}{\ensuremath{\mathit{L}_{\star}}}                 
\newcommand{\mdot}{\ensuremath{\dot{M}}}                             
\newcommand{\mstar}{\ensuremath{\mathit{M}_{\star}}}                 
\newcommand{\rstar}{\ensuremath{\mathit{R}_{\star}}}                 
\newcommand{\teff}{\ensuremath{\mathit{T}_{\rm eff}}}                
\newcommand{\reff}{\ensuremath{\mathit{R}_{\rm phot}}}                
\newcommand{\vinf}{\ensuremath{\upsilon_{\infty}}}                          
\newcommand{\tstar}{\ensuremath{\mathit{T}_{\star}}}                 
\newcommand{\K}{\ensuremath{\mathrm{K}}}                 

\newcommand{\vrot}{\ensuremath{\upsilon_{\rm rot}}}                         
\newcommand{\vcrit}{\ensuremath{\upsilon_{\rm crit}}}                         

\newcommand{\lam}{\ensuremath{\lambda}}                 

\newcommand{\magu}{\ensuremath{\mathit{M}_{\it U}}}

\newcommand{\magr}{\ensuremath{\mathit{M}_{\it R}}}    
\newcommand{\magi}{\ensuremath{\mathit{M}_{\it I}}}

\newcommand{\mbol}{\ensuremath{\mathit{M}_{\rm bol}}}

\begin{document}

\title{Fundamental properties of core-collapse Supernova and GRB progenitors: predicting the look of  massive stars before death}

\author{Jose H. Groh\inst{1},  Georges Meynet\inst{1},   Cyril Georgy\inst{2}, and Sylvia Ekstr\"om\inst{1}}

\institute{
Geneva Observatory, Geneva University, Chemin des Maillettes 51, CH-1290 Sauverny, Switzerland; \email{jose.groh@unige.ch}
\and 
Astrophysics group, EPSAM, Keele University, Lennard-Jones Labs, Keele, ST5 5BG, UK
}
\authorrunning{Groh et al.}
\titlerunning{Fundamental properties of core-collapse SN progenitors }

\date{Received  / Accepted }

\abstract{We investigate the fundamental properties of core-collapse Supernova (SN)  progenitors from single stars at solar metallicity. For this purpose, we combine Geneva stellar evolutionary models with initial masses of $\mini=20-120~\msun$ with atmospheric/wind models using the radiative transfer code CMFGEN. We provide synthetic photometry and high-resolution spectra of hot stars at the pre-SN stage.  For models with  $\mini=9- 20~\msun$, we supplement our analysis using publicly available MARCS model atmospheres of RSGs to estimate their synthetic photometry. We employ well-established observational criteria of spectroscopic classification and find that massive stars, depending on their initial mass and rotation, end their lives as red supergiants (RSG), yellow hypergiants (YHG), luminous blue variables (LBV), and Wolf-Rayet (WR) stars of the WN and WO spectral types. For rotating models, we obtained the following types of SN progenitors: WO1--3  ($\mini \geq 32~\msun$), WN10--11 ($25 < \mini < 32~\msun$), LBV  ($20 \leq \mini \leq 25~\msun$), G1 Ia$^+$ ($18 < \mini < 20~\msun$), and  RSGs  ($9 \leq \mini \leq 18~\msun$). For non-rotating models, we found spectral types WO1--3  ($\mini > 40~\msun$), WN7--8 ($25 < \mini \leq 40~\msun$), WN11h/LBV  ($20 < \mini \leq 25~\msun$), and RSGs ($9 \leq \mini \leq 20~\msun$). Our rotating models indicate that SN IIP progenitors are all RSG, SN IIL/b progenitors are 56\% LBVs and 44\% YHGs, SN Ib progenitors are 96\% WN10-11 and 4\% WOs, and SN Ic progenitors are all WO stars. We find that not necessarily the most massive and luminous SN progenitors are the brighter ones in a given filter, since this depends on their luminosity, temperature, wind density, and how the spectral energy distribution compares to a filter bandpass. We find that SN IIP progenitors (RSGs) are bright in the $RIJHK_S$ filters and faint in the $UB$ filters. SN IIL/b progenitors (LBVs and YHGs), and SN Ib progenitors (WNs) are relatively bright in optical/infrared filters, while SN Ic progenitors (WOs) are faint in all optical filters. We argue that SN Ib and Ic progenitors from single stars should be undetectable in the available pre-explosion images with the current magnitude limits, in agreement with observational results. 
}

\maketitle

\section{\label{intro}Introduction} 

Massive stars are ubiquitously present in the local and far Universe. Due to their short lives and the physics of star formation, they are easily outnumbered by their low-mass siblings like the Sun. Nevertheless, the impact of massive stars through cosmic time is far from negligible, as they are the main responsible for the input of ionizing photons, energy, and recycling mass into the interstellar medium through stellar winds and Supernova (SN) explosions.

Mass loss and angular momentum evolution  play a key role in determining the properties of massive stars across their evolution and the properties of the SN progenitor and the ensuing explosion \citep[for recent reviews see][]{mm12,langer12}. The current generation of Geneva evolutionary models \citep{ekstrom12,georgy12a} predict that single rotating stars with initial masses (\mini) in the range $8~\msun \lesssim \mini \lesssim 17~\msun$ end their lives as red supergiant (RSG) stars before a SN event of the type IIP (i.e., with H lines dominating the spectrum and a plateau in the lightcurve). This scenario is well supported by the observations of SN IIP progenitors in pre-explosion images, which have been shown to be RSGs with $8.5~\msun \lesssim \mini \lesssim 16.5~\msun$ \citep{smartt09b}.

The agreement between theory and observations of SN progenitors is much less satisfactory for stars with  $\mini \gtrsim 17~\msun$. One problem is related to the fact that Galactic RSGs are observed to evolve from stars with $\mini$ up to 25-30~\msun\ \citep{levesque05}. If these stars die as RSGs, this raises  the issue of why no RSG more massive than about 16.5~\msun\ has been detected in pre-explosion images of  SN progenitors (``the red supergiant problem", \citealt{smartt09a}). A possible solution to this problem would be the presence of circumstellar extinction around RSGs, which would underestimate the luminosity and mass determinations 
from the pre-explosion images \citep{smith11sn,walmswell12}. It may also be that the most massive stars evolve away
from the RSG phase and end their lifetime in a bluer portion of the HR diagram \citep{vanbeveren98a,salasnich99,yooncantiello10,georgy12a,georgy12,meynet13}. For instance, rotating models predict that stars with $20~\msun \lesssim \mini \lesssim 25 ~\msun$ are born as O dwarfs 
and exhibit a spectrum reminiscent of the rare luminous blue variable (LBV) stars 
before the SN explosion \citep{gme13}. The models indicate that a small amount of H is present in the 
envelope (a few $10^{-2}~\msun$), making it difficult to infer the kind of SN that will follow core collapse 
(SN IIL, SN IIb, or even SN IIn if significant circumstellar material surrounds the progenitor). 

The situation becomes even hazier for stars with $ \mini \gtrsim 25~\msun$. According to evolutionary models, they reach the Zero-Age Main Sequence as O-type stars burning hydrogen in their cores and evolve to Wolf-Rayet (WR) type, helium-core burning stars (see e.g., \citealt{maeder_araa00,meynet03,langer07}). Between the O-type and WR stages, the star may or may not go through an unstable, short-lived stage, usually associated to the LBV phase, and/or through a RSG phase. The models predict that core-collapse SN occur after the WR phase, and are of the type Ibc \citep{georgy12a,georgy09}.  However, this scenario has yet to be observationally confirmed, specially because WR stars have not been directly detected as SN Ibc progenitors yet \citep{smartt09a,eldridge13}. The non-detection could be due to the intrinsic faintness of WRs in the optical bands, which are the ones usually available for the SN explosion site \citep{yoon12}, or because SN Ibc progenitors have lower masses and result from binary evolution \citep{pod92,smartt09a,eldridge13}, or both. In addition, observations suggest that some massive stars can explode as SNe already during the LBV phase \citep[e.g.][]{kv06,smith07,pastorello07,galyam09}. Some of these progenitors likely had $ \mini \gtrsim 50~\msun$, which is much higher than the range of LBVs that can be SN progenitors based on single stellar evolution models ($20~\msun \lesssim \mini \lesssim 25 ~\msun$; \citealt{gme13}). 

Such glaring discrepancies between observations and the stellar evolution theory exposes a main gap in the understanding of massive stars and highlights our incomplete view of their post-Main Sequence evolution and fate. Several reasons exist for explaining our limit knowledge of massive stars. For instance, binarity plays an important role in the evolution of massive stars, as a significant fraction of massive stars seems to be in systems that will interact \citep{sana12}. Neglecting the effects of binaries on the properties of an average population of massive stars would likely yield inconsistent results. An equally important reason for our limited knowledge concerns the challenging comparison between observed data and stellar evolution models. Stellar evolution models are able to predict the stellar parameters only up to the stellar hydrostatic surface, which is not directly comparable to the observations when a dense stellar wind is present. This is the case for massive stars, in particular at their post-Main Sequence evolution, when they lose mass at enormous rates ($10^{-6}$ to $10^{-3}~\msunyr$). When the stellar wind becomes dense, eventually the photosphere becomes extended and is formed in a moving layer. Numerous emission lines arise in the wind, veiling (sometimes completely) the underlying spectrum from the hydrostatic surface that would be observable otherwise. As a consequence, the outputs of massive star evolution models, such as temperature, luminosity, and surface abundances at the hydrostatic surface, are difficult to be directly compared to observed quantities, such as a spectrum or a photometric measurement.

To solve this issue,  it is necessary to couple stellar evolutionary calculations to radiative transfer models of the stellar atmosphere and wind. In \citet{gme13} we presented, for the first time, combined stellar evolution and atmospheric modeling at the pre-SN stage, for stars with $\mini=20$ and 25~\msun. Surprisingly, we found that these stars end their lives remarkably similar to LBVs, showing that in a {\it single stellar evolution framework, massive stars can explode as LBVs.} 

Here we present a theoretical investigation of core-collapse SN progenitors from single stars with initial masses in the range 9--120~\msun\ at solar metallicity. We analyze how they appear to a distant observer, predicting observables such as the high-resolution spectrum, absolute magnitudes, colors, and bolometric corrections. The comparison between observations of SN progenitors and models can shed light on several properties of the progenitor, such as its mass, chemical composition, and initial rotation rate. 

The motivations for our investigation are numerous. First, it allows a better constrain of the stellar evolution models. The classification of SNe generally traces the presence of chemical elements in the spectrum and the shape of the lightcurve (see e.g., \citealt{fillipenko97} for a review). Studying the relative rate of the different core-collapse SN types is an important constraint for stellar evolution. This is because the chemical abundances in the ejecta supposedly reflects those from the progenitor before the SN explosion. Second, it allows to link a given observed SN rate to a given star formation history. This is only possible if we know sufficiently well the nature of the progenitors for the different types of core-collapse SNe. Finally, producing realistic observables from single-star evolution models also allows one to properly gauge whether a binary evolution scenario must necessarily be invoked to explain the observed properties of a given SN progenitor.

This paper is organized as follows. In Sect.~\ref{model} we describe our modeling approach. In Sect. \ref{ccsnfrac} we revisit the expected fraction of the different kinds of core-collapse SNe  from single stars and how they compare to recent observations. In Sect.~\ref{teffhr} the location of the SN progenitors on the HR diagram, luminosity, and effective temperature are presented.  Their spectroscopic appearance and spectral classification are performed in Sect.~\ref{spec}, while Sect. \ref{specchem} compares our results with previous classifications of spectral types of SN progenitors based on chemical abundance criteria. The absolute magnitudes and bolometric corrections as a function of initial mass of the progenitor are presented in Sect. \ref{absmag}. We investigate the detectability of progenitors of the different types of core-collapse SN in Sect. \ref{detect}.  We discuss the surprising finding that WO stars are progenitor of type Ibc SNe in Sect. \ref{wosn}, while in Sect. \ref{lbvssn} we discuss the possibility that LBVs are SN progenitors. We present our concluding remarks in Sect. \ref{conc}. 

In a series of forthcoming papers, we will present the results for the complete evolution of massive stars and for a larger metallicity range, and investigate the effects of several physical ingredients, such as magnetic fields, on the final appearance of massive stars.

\section{\label{model}Physics of the models}

\subsection{Stellar evolution}

The evolutionary models are computed with the Geneva stellar evolution code. The majority of the models are those from \citet{ekstrom12}. To better determine the mass ranges of the different SN progenitors,  we compute new models with \mini\ of 16.5, 18, and 28~\msun\ (rotating models), and 23 and 50~\msun\ (non-rotating models). All models are publicly available through the webpage \url{http://obswww.unige.ch/Recherche/evol/-Database-}. Here we summarize the main characteristics of the code, and refer the reader to the aforementioned papers for further details. The models assume solar metallicity (Z=0.014), initial abundances from \citet{asplund09}, and the rotating models have initial rotational speed (\vrot) of 40\% of the critical velocity (\vcrit). The prescription for the rotational diffusion coefficients is taken from \citet{zahn92} and \citet{maeder97}. 

Mass loss is a key ingredient of the models, affecting not only the final position in the HR diagram but also the emerging spectrum of hot stars. Since the stellar evolution code requires previous knowledge of the spectral types to adopt a certain mass-loss recipe, criteria based on chemical abundances and effective temperatures are employed to estimate the type of massive star (OB, WR, RSG) at each time step \citep{smithmaeder91,meynet03}. The radiative mass-loss rates for OB stars follow the \citet{vink01} prescription, while for WR stars the \citet{nl00} and \citet{grafener08} prescriptions are employed. For the RSGs, when $\log (\teff/\K) > 3.7$, the \citet{dejager88} prescription is applied for initial masses of 15~\msun\ and above. For  $\log (\teff/\K) \leq 3.7$, a linear fit to the data from \citet{sylvester98} and \citet{vanloon99} is applied (see also \citealt{crowther01}). For the RSGs in models below 15~\msun, the \citet{reimers75,reimers77} relation (with $\eta=0.5$) is used. 

Because of variations in the ionization level of hydrogen beneath the surface of the star during the RSG phase, significant changes in opacity may occur. Thus, some layers might exceed the Eddington limit, possibly driving instabilities. In this case, the radiative mass loss is increased by a factor of three in our models with initial mass above 18~\msun, which matches the \mdot\ determinations from \citet{vanloon05}. The effects of different \mdot\ recipes during the RSG phase on the evolution of massive stars has been investigated by \citet{georgy12}, to where we refer the interested reader for further details.

As in \citet{ekstrom12}, the stellar evolution models employed here terminate at the end of core-carbon burning. We do not expect significant variations in the surface properties after this phase. To verify this assumption, we computed the non-rotating 60~\msun\ model until core O burning, and negligible changes in \lstar\ and \teff\ were seen ($\sim 0.01$ dex).

\subsection{Atmospheric and wind modeling of hot stars }

The model spectra computed here are publicly available through the webpage \url{http://obswww.unige.ch/Recherche/evol/-Database-}.

To compute the output spectra of stars with $\tstar > 8000~\K$ we used the atmospheric radiative transfer code CMFGEN \citep{hm98}. CMFGEN is a  spherically-symmetric, fully line blanketed code that computes line and continuum formation in non-local thermodynamical equilibrium. Since all  evolutionary models discussed here present negligible surface rotation at the pre-SN stage, the use of spherical symmetry is well justified. CMFGEN computes a self-consistent radiative transfer including the stellar hydrostatic surface and the wind. Wind (micro) clumping is included with a volume filling factor ($f$) approach, which assumes dense clumps and a void interclump medium. The wind is also assumed to be unclumped close to the stellar surface and to acquire full clumpiness at large radii. All models computed here assume $f=0.1$. CMFGEN does not solve the momentum equation of the wind, and thus a hydrodynamical structure must be adopted. For the wind part, we assume a standard $\beta$-type law with $\beta=1$, while a hydrostatic solution is computed for the subsonic portion. This is applied up to 0.75 of the sonic speed, where the hydrostatic and wind solutions are merged. The wind terminal velocity ($\vinf$) is computed using the parametrization from \citet{kudritzki00} for OB stars and LBVs, and from \citet{nl00} for WR stars of the WN and WC type. For WO stars, an iterative scheme is adopted. We initially compute a spectrum with the value of $\vinf$ as given by the \citet{nl00} recipe, which is typically at most $\sim2800~\kms$. If a WO-type spectrum arises, we recompute a spectrum with $\vinf=5000$~\kms\, which is more representative of the observed Galactic WO stars \citep{drew04,sander12}.

We use the outputs from the stellar structure calculations with the Geneva code, such as the radius, luminosity, mass, and surface abundances, as inputs in CMFGEN. For consistency, we adopt the same mass-loss rate recipe as that used by the Geneva evolution code. We use the temperature structure of the stellar envelope to merge the CMFGEN solution and the stellar structure solution. The outputs from the CMFGEN calculations that we discuss here are the synthetic spectrum, photometry, and the effective temperature \teff, defined as the temperature of the layer where the Rosseland optical depth is 2/3. The values of $\tstar$ quoted here correspond to those predicted by the Geneva stellar evolution code without the correction due to the optical depth of the wind, and {\it not} to the temperature at a fixed Rosseland optical depth (usually 20).

\subsection{Atmospheric modeling of cool stars}

A realistic atmospheric analysis of luminous cool stars requires the inclusion of convection and H$^-$ and molecular opacities, which at the moment are not included in CMFGEN.  Here, we employ the publicly available MARCS models \citep{gustafsson08} to perform synthetic photometry of the SN progenitors that have $\teff=3400-5400~\K$ and thus are RSGs or YHGs. We use the model grids that have abundances corresponding to CN-processed material, which is characteristic of RSGs at the pre-SN stage. A mass of 5~\msun\ and turbulent velocity of $2~\kms$ are assumed. Since these MARCS models are available only at coarse $\teff$ sampling (3400, 3600, and 3800  K for RSGs; 5000, 5250, 5500, 5750 K for YSG/YHG) and not at the exact luminosities predicted by the evolutionary models, the magnitudes and bolometric corrections of the cool SN progenitors were estimated by:\\
1) linearly interpolating the magnitudes (bolometric corrections) of two bracketing MARCS models in $\log \teff$ space to the desired $\teff$ value predicted by the Geneva code. The bracketing MARCS models were previously scaled to the same luminosity;\\
2) scaling the interpolated magnitudes to the luminosities predicted by the Geneva code.\\
This is a zeroth-order approximation to estimate the magnitudes and should be checked in the future against MARCS models computed specifically for the physical parameters ($\lstar$, \teff, \mstar, abundances) found at the pre-SN stage. Since these are unavailable at the moment, synthetic spectra of RSGs and YHGs are not provided. Still, the magnitude estimates computed here provide important insights into the nature of core-collapse SN progenitors.

\section{The core-collapse SN fraction from single stars}
\label{ccsnfrac}

\defcitealias{smith11sn}{S11}
\defcitealias{eldridge13}{E13}

Before computing the stellar spectra and colors associated to the various SN progenitors predicted by the grids of 
models of \citet{ekstrom12}, it is interesting to see whether these models can reproduce 
or not the observed rates of various types of core-collapse SNe. 
This has already been discussed extensively in \citet{georgy12a}, which determined the relative rates of the different core-collapse SN types and compared to the observational sample from \citet{bp09}. Here we revisit this topic, given that two relatively large samples of observational core-collapse SN data have been recently released (\citealt{smith11sn}, hereafter S11, and \citealt{eldridge13}, hereafter E13).  We assume, as in \citet{georgy12a}, a Salpeter-type initial mass function, that the lower and upper initial mass limit for core-collapse SNe are 8 and 120~\msun, respectively, that SNe occur even when black holes are formed, and that all SNe are observable (see also \citealt{heger03}). For the SN classification  we assume, as in \citet{georgy12a}, the following chemical abundances in the ejecta:\\
-- a SN IIP has more than 2~\msun\ of H; \\
-- a SN IIL/b has between 0 and 2~\msun\ of H;\\
-- a SN Ib has no H and more than 0.6~\msun\ of He;\\
-- a SN Ic has no H and less than 0.6~\msun\ of He.

We present in Table \ref{snrates} the fraction of the different SN types predicted by our non-rotating and rotating models (see also \citealt{georgy12a}), and those of the S11 and E13 observational samples. A detailed comparison between both observational samples has been performed by \citetalias{eldridge13}.

\begin{table}
\caption{Core-collapse SN fractions from our models compared to observed samples}
\footnotesize
\label{snrates}
\begin{tabular}{lccccc}
\hline
\hline         
SN Type 		& Non-rot. model & Rot. model  & Obs. E13 & Obs. S11\\
                             &    (\%)                        &    (\%)                        &    (\%)                        &    (\%)                             \\                         
\hline
\smallskip
IIP     	          	& 70.7                  &     64.9                         &    55.5 	$\pm$ 6.6      &  48.2 $^{+5.7}_{-5.6}$  \\		
\smallskip 
IIL + IIb    		         & 16.2                  &     13.1                          &     15.1	$\pm$ 3.4      &  17.0 $^{+4.6}_{-4.0}$ \\ 
\smallskip
IIn      		         & $-$                       & $-$                                &      2.4 	$\pm$ 1.4      &  8.8 $^{+3.3}_{-2.9}$  \\
\smallskip
IIpec                             & $-$                       & $-$                                 &      1.0 	$\pm$ 0.9      &  $-$  \\
\smallskip
Ib      			         & 8.1                     &     7.8                            &        9.0 	$\pm$ 2.7      &  8.4 $^{+3.1}_{-2.6}$  \\
\smallskip
Ic      			         & 5.0                     &     14.2                         &         17.0 	$\pm$ 3.7       & 17.6 $^{+4.2}_{-3.8}$  \\
\hline
\hline
\end{tabular}
\end{table}

As one can see, the SN rates yielded by our rotating models are mostly within the errors of the S11 and E13 observations, 
with the exception of the SN IIP fraction, which is $1.5\sigma~(2.9\sigma)$ above the E13 (S11) values. 
For the non-rotating models, the SN IIP relative rate is too high compared to the observed one, 
while the SN Ic relative rate is too low. We stress that the SN rates are quite sensitive to metallicity \citep[e.g,][]{georgy09} and, while our models discussed here are at solar metallicity, the observed samples likely include a range of metallicities around solar.

\begin{figure*}
\centering
\resizebox{0.497\hsize}{!}{\includegraphics{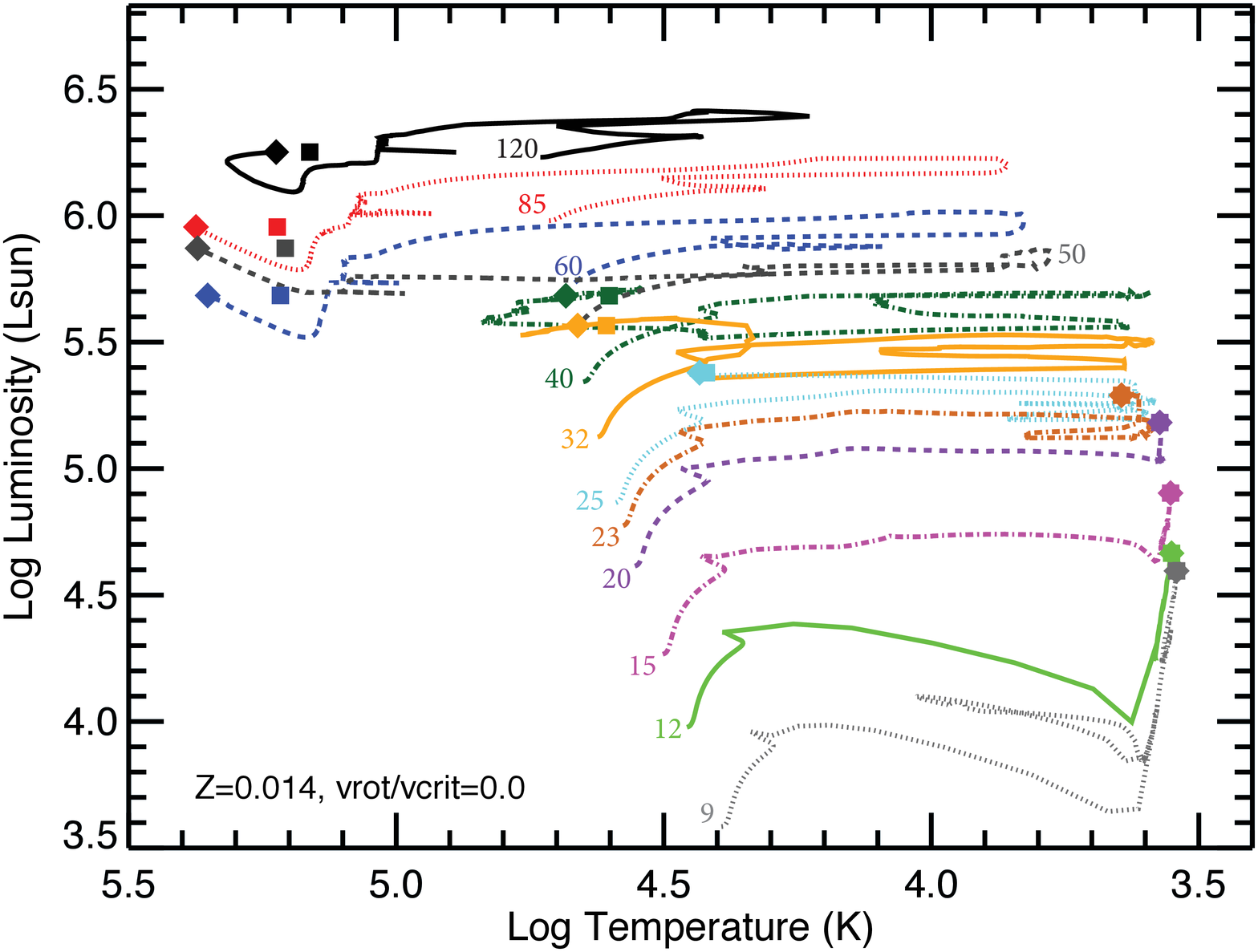}}
\resizebox{0.497\hsize}{!}{\includegraphics{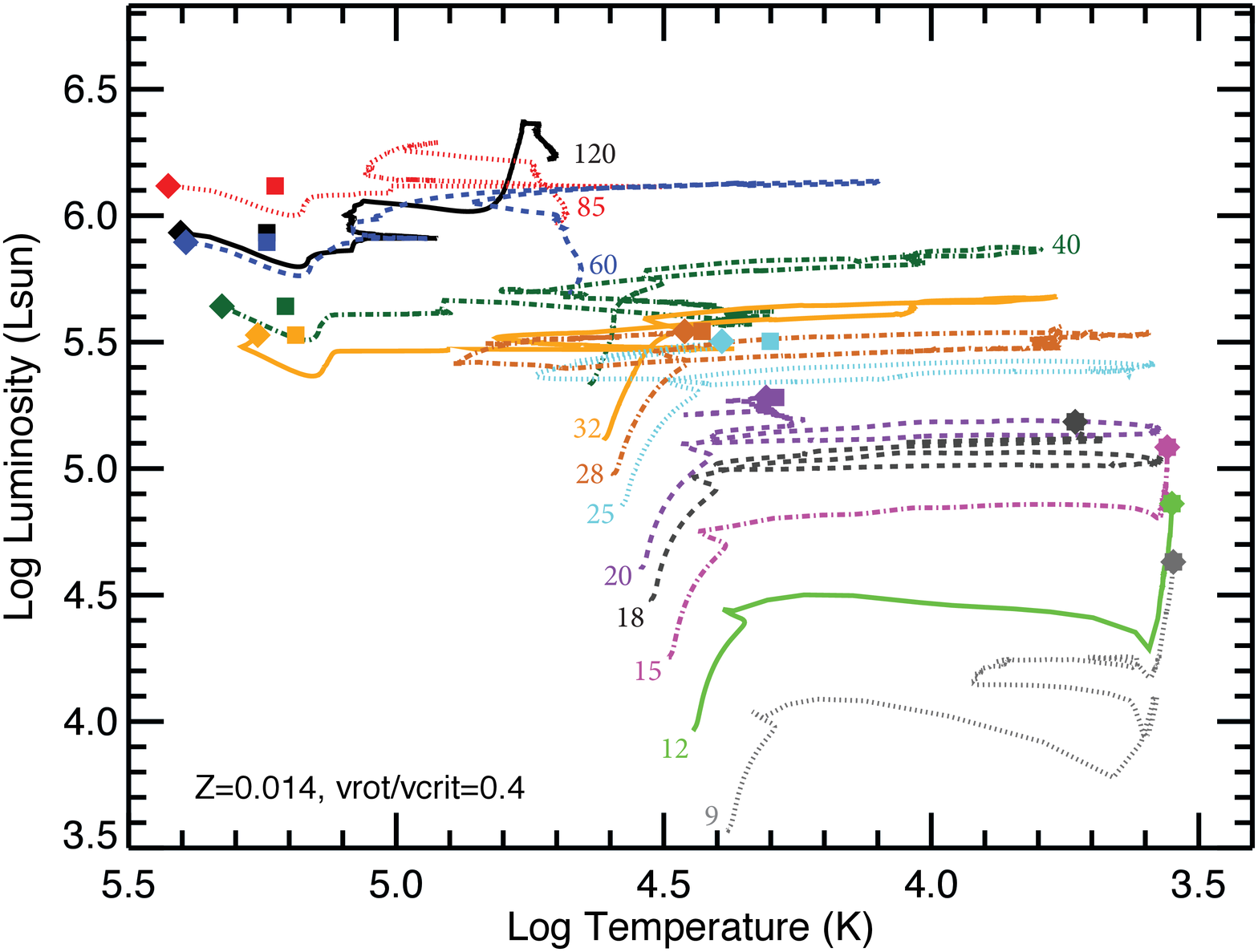}}
\caption{\label{hrd1}{ {\it (Left)}: HR diagram showing evolutionary tracks of non-rotating stars with initial masses between 9~\msun\ and 120~\msun\ at solar metallicity (Z=0.014). The diamonds correspond to the temperature computed by the Geneva code and are not corrected by the effects of the stellar wind ($\tstar$), and the squares correspond to the values of $\teff$ computed either with CMFGEN (for $\teff > 8000$ K), or assumed to be equal to $\tstar$ for RSGs (causing the squares and diamonds to overlap). {\it (Right)}: Similar to the left panel, but for rotating stars with initial $\vrot/\vcrit=0.4$.  Except for the non-rotating 23 and 50~\msun\ and the rotating 18 and 28~\msun\ models, evolutionary tracks are from \citet{ekstrom12}.}}
\end{figure*}

Interesting remarks can be deduced from these results. First,
from a pure theoretical point of view, the different results obtained with and without rotation illustrate
the sensitivity of the predicted supernova rate on axial rotation. Second, for
a proper comparison with the observations, population synthesis models are needed, 
taking into account a range of initial rotational velocities (our rotating models have initial \vrot/\vcrit=0.4).
Third, the current rotating models are in good shape for reproducing the observed rates, while non-rotating
models are in a less good position. This is in line with other evidences for some rotational mixing
being active in massive stars (see e.g. the review by Maeder \& Meynet 2012 and references therein).
Fourth, the above comparisons somewhat support the current rotating models.
However, it is difficult to draw very firm conclusions here since
we cannot discard the possibility that some of our models would produce some faint or even failed
SN event. In that sense, the theoretical SN Ibc rates are likely overestimated. If this turns out to occur, it may indicate that other channels are important  for explaining the observed rates, as for instance the close binary evolution channel (\citetalias{smith11sn}, \citetalias{eldridge13}). 

Since a significant fraction of massive stars are in binary systems \citep[e.g.][]{sana12}, it would not be surprising if a sizable fraction of core-collapse SN progenitors evolve under the influence of a close companion, for instance via tides and mass transfer \citep{demink13}. The modeling of SN Ibc lightcurves also provides strong constraints to the amount of ejecta mass \citep[e.g.][]{drout11,dessart11b,cano13}, with an average value of 4.6~\msun \citep{cano13}. Since our models predict SN Ibc ejecta masses in the range 6.7-- 21.8~\msun\ \citep{georgy12a}, it is thought that part of the SN Ibc progenitors could come from binary evolution. Based on the core-collapse SN rates, both \citetalias{smith11sn} and \citetalias{eldridge13} independently find that roughly 50\% of the progenitors follow single star evolution and 50\% evolve in interacting binary systems.

Ultimately, the best way to distinguish between SN progenitors that come from single or binary evolution is the direct detection of the progenitor in pre-explosion images. To fully grasp all the information from these observations, a comparison with stellar evolution models is needed. The main driver of this paper is thus to produce observables out of single massive star evolution models at the pre-SN stage that can be directly compared to observations.

\begin{table*}
\begin{minipage}{\textwidth}
\caption{Properties of the core-collapse SN progenitor models at solar metallicity. From left to right, the columns correspond to the initial mass on the Main Sequence (\mini), the SN progenitor mass (M$_\mathrm{prog.}$), and the age, bolometric luminosity (\lstar), temperature at the hydrostatic surface (\tstar), effective temperature (\teff), radius of the hydrostatic surface (\rstar), photospheric radius (\reff), mass-loss rate (\mdot), wind terminal velocity (\vinf), steepness of the velocity law ($\beta$),  and spectral type at the pre-SN stage. The last column corresponds to the SN type as computed by \citet{georgy12a}.
 Spectral types in italic were not obtained via classification of the spectra, and thus are only tentative.}
\label{model log}
\footnotesize
\centering
\vspace{0.1cm}
\begin{tabular}{r c c c c c c c c c c c c}
\hline\hline
M$_\mathrm{ini}$& M$_\mathrm{prog.}\tablefootmark{a} $& Age\tablefootmark{a}  & $\lstar$\tablefootmark{a}  & $\tstar$\tablefootmark{a,b}  & $\teff$ & $\rstar$ & $\reff$ & $\mdot$\tablefootmark{a}  & $\vinf$ & $\beta$ & Sp. Type  & SN Type \\
(\msun) & (\msun) &  (Myr) & (\lsun) & (K) & (K) & (\rsun) & (\rsun) & (\msunyr) & (\kms) & & \\  
\hline
\multicolumn{13}{c}{non-rotating models }\\
\hline                                                                                                     					   
  9 & 8.8 & 30.3 & 39401 & 3480 & 3480 & 547.49 & 547.49 & 9.7$\times 10^{-7}$ & $-$ & $-$ & $K-M ~I$                    		& IIP \\	 	  
12 &  11.3  & 17.8 & 46259 & 3556 & 3556 & 567.94 & 567.94 &  1.3$\times 10^{-6}$  & $-$ & $-$ & $K-M ~I$ 						  &  IIP  \\	  
15 & 13.25 & 12.5 & 79999 & 3571 & 3571 & 740.91  & 740.91  & 3.2$\times 10^{-6}$ & $-$ & $-$ & $K-M ~I $ 					    &  IIP \\		
20 &   8.8  &   8.7 & 152291 & 3741  & 3741 &931.42 & 931.42   & 2.0$\times 10^{-5}$  & $-$ &$-$ &$ K-M~ I $						     &  IIL/b  \\
23 &   7.6  &   7.6 & 195030 & 4410  & 4410 &758.29 & 758.29   & 4.4$\times 10^{-5}$  & $-$ &$-$ &$ K-M ~I $
 		& IIL/b\\     
 25  & 8.2 & 7.09 & 2.4E+05 &  27116 &  26330 & 22.20 & 23.49 & 9.7$\times 10^{-6}$ &   694 & 1.0 & WN11h/LBV  						&    IIL/b \\   
 32  & 10.9 & 5.81 & 3.7E+05 &  45790 &  40480 &  9.65 & 12.32 & 2.9$\times 10^{-5}$ &  1200 &   1.0  & WN7--8o					        &   Ib\\	
 40 & 12.7 &  4.97  & 4.8E+05 &  48183 &  40006 & 9.99  & 14.4 & 4.2$\times 10^{-5}$ & 1291 & 1.0  & WN7--8o 					       &  Ib \\		
 50 & 16.53 & 4.34 & 7.4E+05 & 235060 & 161200 & 0.52 & 1.10 & 1.2$\times 10^{-5}$ & 5000 & 1.0 & WO1--3							& Ib \\			
 60 & 12.4 &  3.97 & 4.8E+05 & 225077 & 164600 &  0.46 &  0.85 & 1.5$\times 10^{-5}$ &  5000 &   1.0  & WO1--3						&  Ic \\		
 85 & 18.5 &  3.40 & 9.0E+05 & 236900 & 166900 &  0.56 &  1.13 & 2.2$\times 10^{-5}$ &  5000 &   1.0  & WO1--3						 &  Ic \\		
 120 & 30.7 & 3.00 &1.8E+06 & 167700 & 145100 &  1.59 &  2.11 & 3.1$\times 10^{-5}$ &  5000 &   1.0  & WO1--3						 &  Ib\\      
 \hline
 \multicolumn{13}{c}{rotating models,  initial $\vrot/\vcrit=0.4 $ }\\
 \hline
    9  &  8.5 & 35.5 &   42802 & 3528 & 3528 & 555.08 & 555.08 &  1.1 $\times 10^{-6}$ & $-$ & $-$ & $  K-M~I$				&  IIP      \\		
   12 & 10.22 & 20.7 & 72652 & 3550 & 3550 & 714.24 & 714.24 & 2.7 $\times 10^{-6}$ & $-$ & $-$ &$  K-M~ I $ 					&  IIP        \\	
   15 & 11.07  & 15.0 & 121442 & 3623 & 3623 & 886.73 & 886.73 & 6.5$\times 10^{-6}$ & $-$ & $-$ & $ K-M~ I$  					 &  IIP  \\			
   16.5 &  12.14 & 13.0  & 126610 &   3645  & 3645 & 894.92 & 894.92 & 7.0$\times 10^{-6}$ & $-$ & $-$ & $  K-M~ I $			 &  IIP  \\			
   18 & 6.32 & 11.7 & 153052 & 5382 & 5382 & 451.20 & 451.20 & 1.7$\times 10^{-5}$ & $-$ & $-$ &$ G1 Ia^+$						&  IIL/b  \\		
  20   & 7.1  & 10.4 &1.9E+05 &  20355 &  19540 & 35.30 & 38.11 & 1.2$\times 10^{-5}$ &   272 & 1.0 & LBV 						&  IIL/b   \\		
  25  & 9.6   & 8.60 & 3.2E+05 &  24625 &  20000 & 31.1 & 46.96 & 4.6$\times 10^{-5}$ &   326 &   1.0 & LBV					 &  Ib    \\		
  28 & 10.8 & 7.92 & 3.5E+05 & 28882 &    26823   & 23.68    &  26.26      & 2.6$\times 10^{-5}$ & 415 & 1.0 & WN10-11			&  Ib    \\			
  32  & 10.1  & 7.22 & 3.4E+05 & 181500 & 154100 &  0.58 &  0.81 & 1.2$\times 10^{-5}$ &  5000 &   1.0 & WO1					&  Ic   \\			
  40 & 12.2  & 6.17  & 4.4E+05 & 211700 & 161100 &  0.49 &  0.85 & 1.4$\times 10^{-5}$ &  5000 &   1.0 & WO1--2					  &  Ic   \\		
 60  & 18.9  & 4.86 &   8.6E+05 & 247100 & 174500 &  0.48 &  1.01 & 2.0$\times 10^{-5}$ &  5000 &   1.0 & WO1--3				  &   Ic \\			
 85 & 26.2   & 4.06 & 1.3E+06 & 266700 & 168400 &  0.54 &  1.34 & 2.8$\times 10^{-5}$ &  5000 &   1.0 & WO1--3					  &  Ib \\			
120  & 18.9 & 3.55 & 8.6E+05 & 252800 & 174500 &  0.48 &  1.01 & 2.0$\times 10^{-5}$ &  5000 &   1.0 & WO1						  &  Ic \\			
 
\hline
\end{tabular}
\tablefoot{\tablefoottext{a}{From \citet{ekstrom12}, except for the non-rotating 23 and 50~\msun\ and the rotating 18 and 28~\msun\ models, which were computed for this work.}\tablefoottext{b}{This is defined here as the temperature computed by the Geneva code without the correction due to optical-depth effects of the stellar wind.}
}
\end{minipage}
\end{table*}

\section{\label{teffhr} Location of SN progenitors in the HR diagram: luminosity and effective temperature}

Figure \ref{hrd1} shows an HR diagram with evolutionary tracks of rotating and non-rotating stars obtained with the Geneva stellar evolution code \citep{ekstrom12,georgy12a},  as well as the additional models computed for the purpose of this work, with initial mass in the range 9 to 120 \msun\  at solar metallicity. Here we focus on the pre-SN stage (diamonds and squares in Fig. \ref{hrd1}).

The temperatures of the evolutionary tracks shown in Fig. \ref{hrd1} refer to those computed by the Geneva code without the correction due to optical-depth effects of the stellar wind. To obtain the effective temperature, optical depth effects due to the presence of an atmosphere and dense stellar wind have to be taken into account, and an atmospheric modeling must be performed. We indicate in Fig. \ref{hrd1} the values of \teff\ obtained with CMFGEN for hot stars (squares). These values thus correspond to the \teff\ that one should employ when comparing with the observed $\teff$. For RSGs, we assume that the presence of a wind does not affect the determination of $\teff$, thus $\teff=\tstar$. The other fundamental parameters of the SN progenitors at solar metallicity are summarized in Table \ref{model log}.

Rotating models with $\sim9~\msun \leq \mini \leq ~16.5\msun$ ($\sim9~\msun \leq \mini \leq ~23\msun$ for non rotating ones) are RSGs at the pre-SN stage \citep{ekstrom12}. They end their lives with $\teff=3480-4410$~K. The rotating 18~\msun\ model is a luminous yellow star at the pre-SN stage, with $\teff=5382~\K$ and $\lstar=1.5\times10^5~\lsun$.

The rotating models with  \mini=20, 25, and 28~\msun\  evolve back to the blue after the RSG phase, and develop strong winds as they evolve towards the end stage. The 20~\msun\ model  achieves $\lstar=1.9\times10^5~\lsun$ and  $\teff\simeq 19\,540$~K, while the 25~\msun\ model has $\lstar=3.2\times10^5~\lsun$ and $\teff\simeq 20\,000$~K at the end of its evolution \citep{gme13}. The rotating 28~\msun\ model reaches $\lstar=3.5\times10^5~\lsun$ and  $\teff\simeq 26\,800$~K at the pre-SN stage. In the non-rotating case, the  $\mini=25~\msun$ model ends its life with $\teff\simeq 26\,300$~K and $\lstar=2.4\times10^5~\lsun$. We note that these models are  close to the bistability limit of line-driven-winds, when abrupt changes in mass loss occur \citep{vink99}.  In some cases, the models flirt with this limit close to the end stages and oscillates from one side of the bistability limit to the other, having an erratic behavior in $\mdot$. Therefore, the final position of these models in the HR diagram are probably relatively coarse within a couple thousand K.

For stars with  $\mini=32~\msun$ to $40~\msun$, there are huge differences in the \teff\ of rotating and non-rotating models at the pre-SN stage. This happens because  rotation increases mixing, which brings He burning processed material (C and O) closer to the surface compared to when rotation is absent. As the star evolves, mass loss removes the outer layers and expose C and O at the stellar surface. Therefore, while the $\mini=32~\msun$ and $40~\msun$ non-rotating models have $\teff\simeq 40\,000$~K, the rotating ones reach $\teff\simeq 150\,000-160\,000$~K at the pre-SN stage. The values of \lstar\ of both rotating and non-rotating models are comparable, with the non-rotating models having a slightly higher luminosity.

We obtain that stars with $\mini=50~\msun$ to 120\msun, both in rotating and non-rotating models, end their lives with extremely high values of  $\teff\simeq145\,000-175\,000$~K.  We notice that the rotating models have a tendency to reach the pre-SN stage with higher \teff\ than the non-rotating models. The difference is about 5\,000 -- 10000 K at 60~\msun\ and 85~\msun, but rises considerably to 30\,000 K at  120~\msun. This occurs because the rotating models have higher \tstar\ than non-rotating models. As discussed by \citet{georgy12a}, the 60~\msun\ and 85~\msun\ rotating models finish with a higher luminosity (0.2 and 0.15 dex, respectively) than the corresponding non-rotating models, while the rotating 120~\msun\ model finishes at a much lower luminosity (by 0.3 dex) than the non-rotating model.

\section{\label{spec} Spectra and spectral type classification of SN progenitors}

While the spectral types of low-mass stars can be securely obtained from $\teff$, $\log g$, and $\lstar$ only, the same is not true for massive stars, in particular at the end of their evolution. This is because massive stars are characterized by dense outflows, which have an impact in the spectral morphology. In addition to the luminosity, temperature, effective gravity, and abundances, several other quantities may affect the emerging spectrum. Among these, the mass-loss rate and wind terminal velocity are the ones that influence most the appearance of SN progenitors. Here, we compute the synthetic spectra of SN progenitors with CMFGEN, based on the ouput from stellar evolution models. Therefore, it is not entirely surprising that the spectral types that we derive in this Section are different from those widely quoted in the literature which, in the absence of a spectrum, were estimated using chemical abundance criteria.

\begin{figure*}
\resizebox{0.995\hsize}{!}{\includegraphics{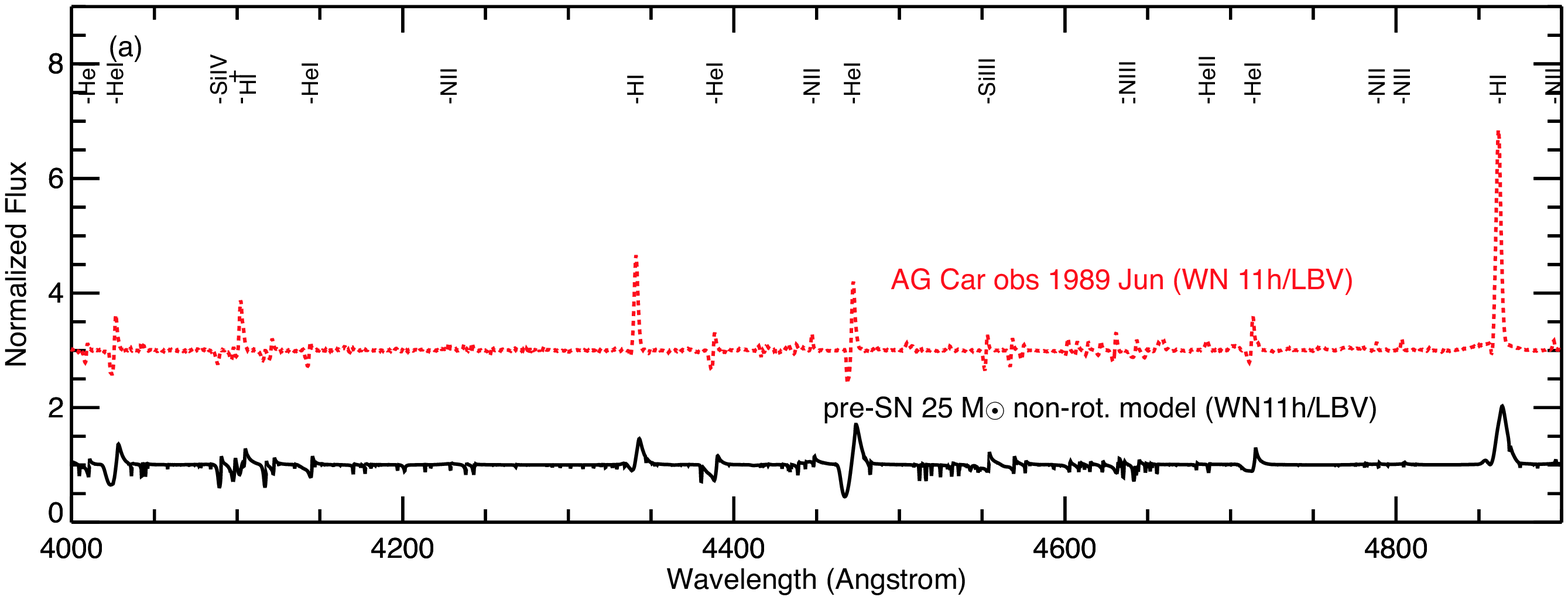}}\\
\resizebox{0.995\hsize}{!}{\includegraphics{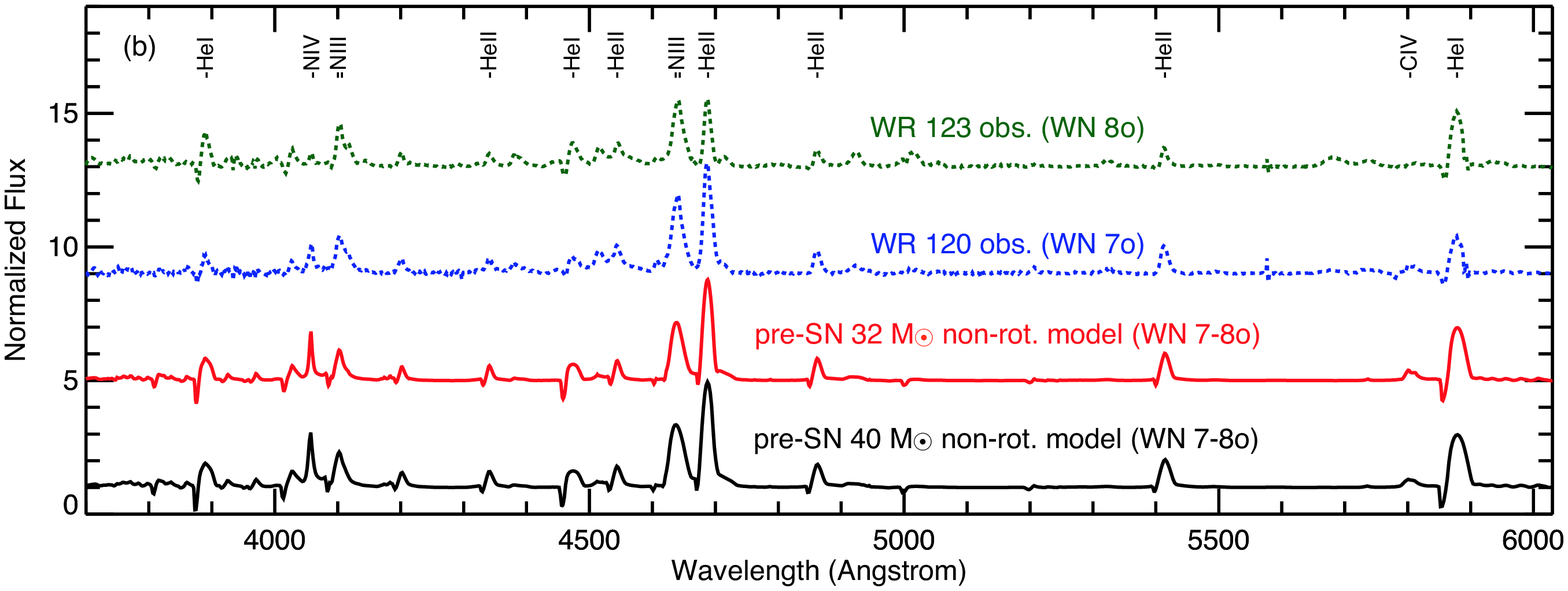}}\\
\resizebox{0.995\hsize}{!}{\includegraphics{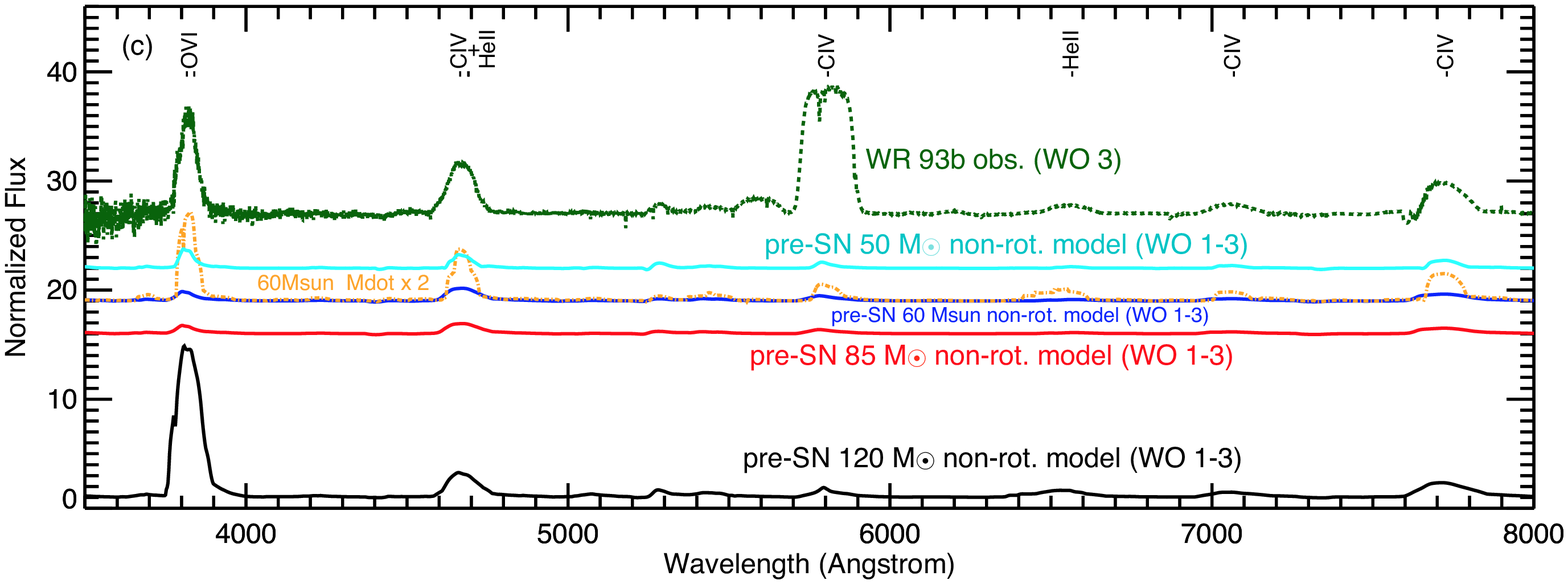}}
\caption{\label{figspec1}{ Montage of the synthetic optical spectra of massive stars at the pre-SN stage from non-rotating stellar evolution models. Observations of stars with similar spectral type (dashed) are shown to support the spectroscopic classifications. The strongest spectral features are indicated. The spectra have been offset in flux for better visualization. {\it (a)}: The 25~\msun\ model (black), which has a WN11h/LBV spectral type. The 1989 June observations of the LBV AG Car (red) are also shown, when it showed a WN11h spectral type \citet{sc94,stahl01,ghd09}.  {\it (b)}: The 32~\msun\ (red) and 40~\msun\ (black) models, which have spectral type WN7--8o, are compared to observations of Galactic WN7o (WR120) and WN8o (WR123) stars, form the catalogue of \citet{hamann95}. {\it (c)}: The 50~\msun\ (cyan), 60~\msun\ (blue), 85~\msun\ (red), and 120~\msun\ (black) models, which have WO1--3 spectral type. The  spectrum of the 60~\msun\ model with $\mdot$ enhanced by a factor of two at the pre-SN stage is shown (orange dot-dashed). The optical spectrum of the Galactic WO 3 star WR 93b (green, from \citealt{drew04}) is also displayed. }}
\end{figure*}

\begin{figure*}
\resizebox{0.995\hsize}{!}{\includegraphics{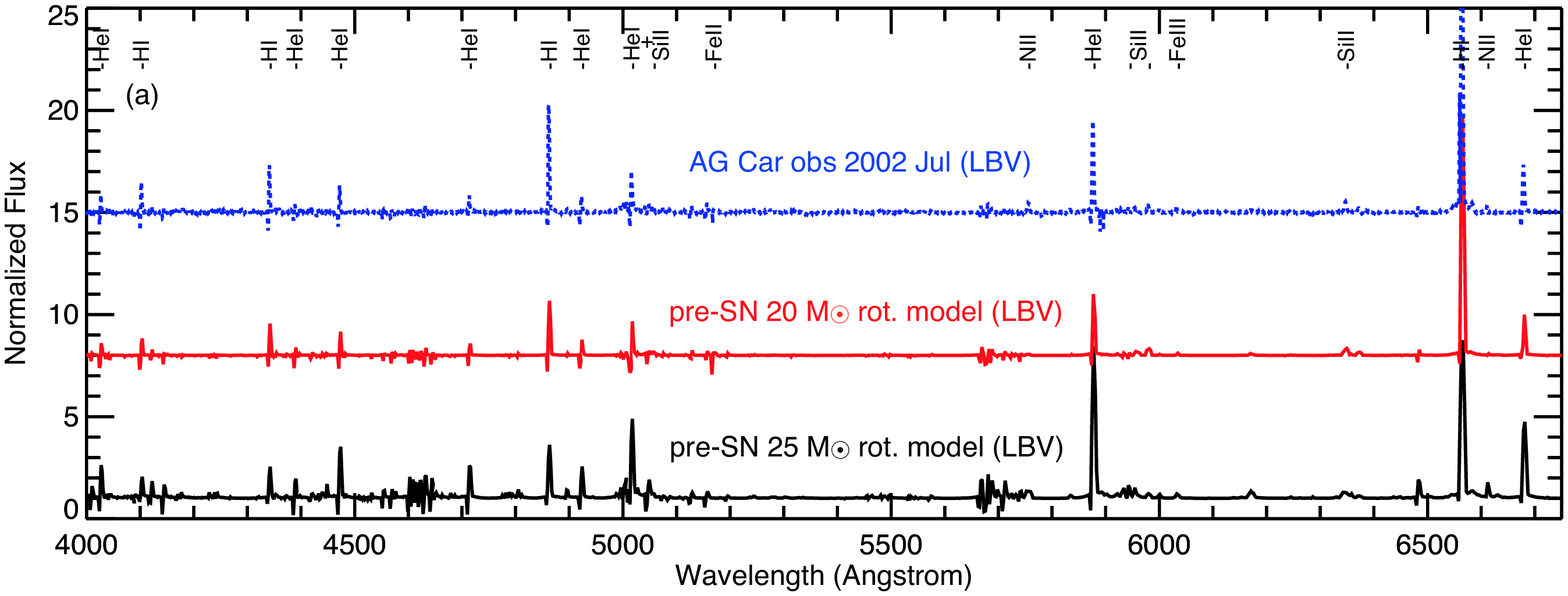}}\\
\resizebox{0.995\hsize}{!}{\includegraphics{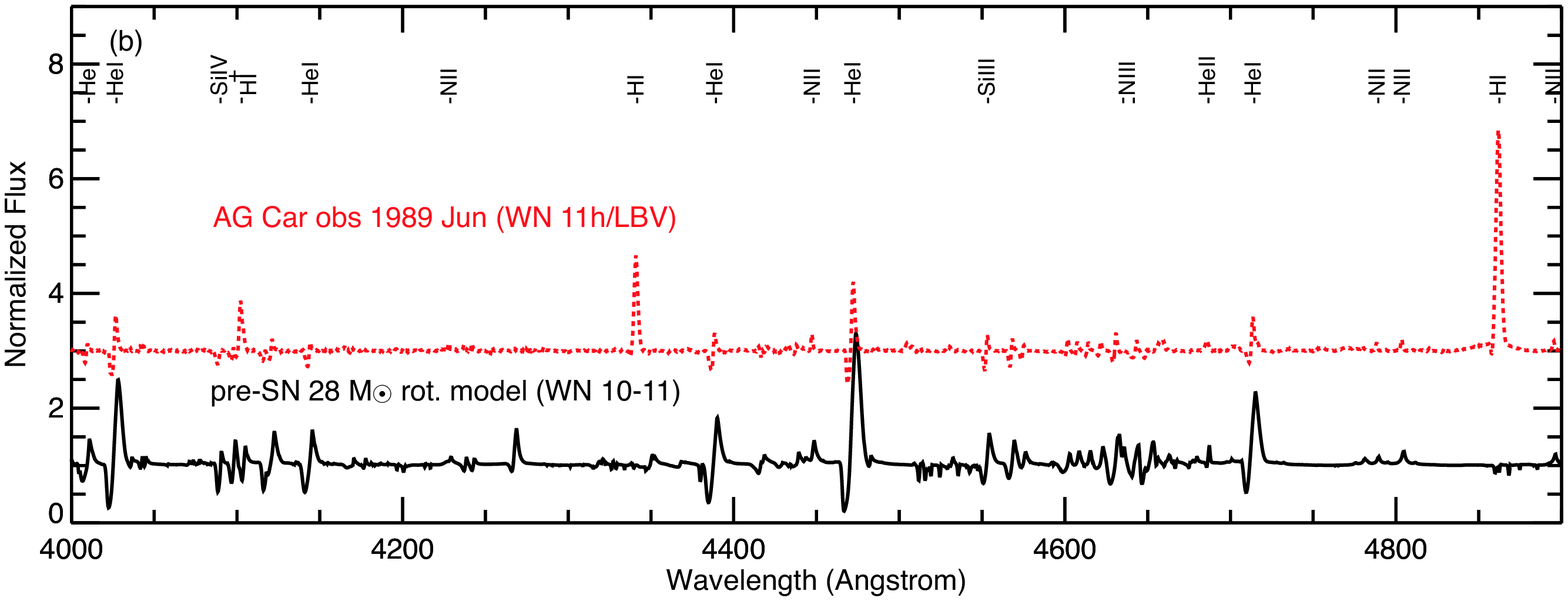}}\\
\resizebox{0.995\hsize}{!}{\includegraphics{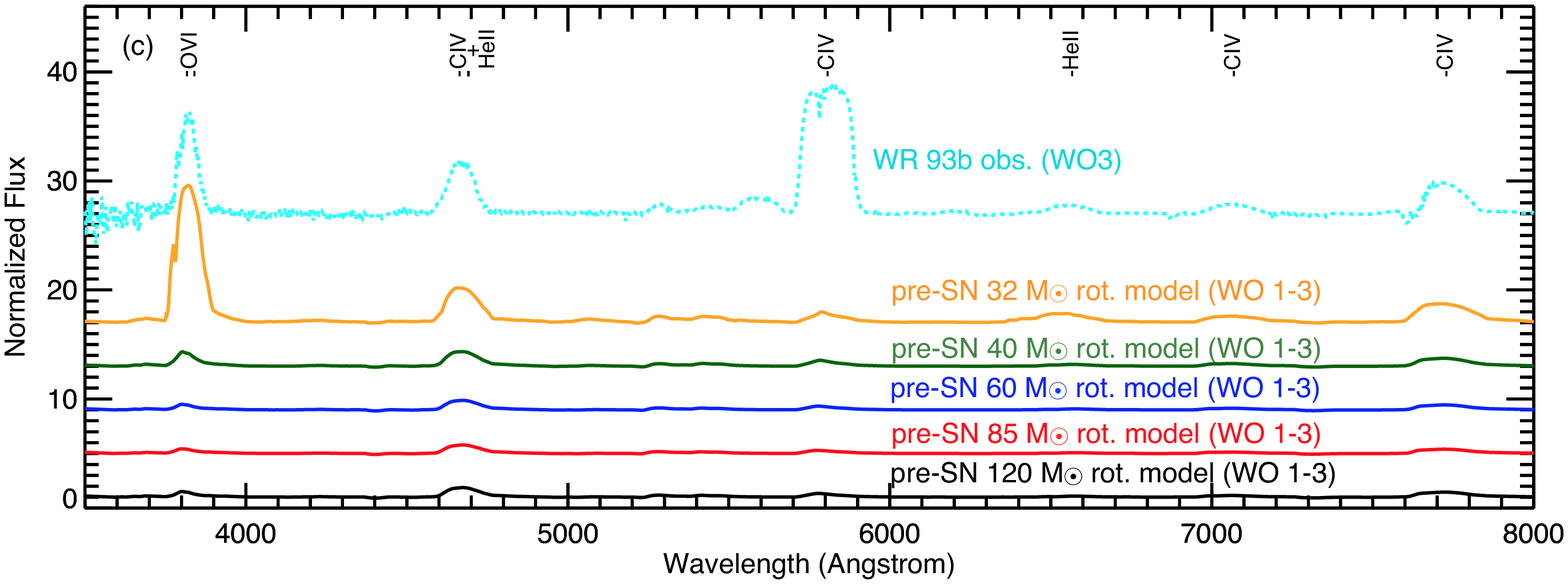}}\\
\caption{\label{figspec2}{Similar to Fig. \ref{figspec1}, but for rotating models at the pre-SN stage. {\it (a)}: The 20~\msun\ (red) and 25~\msun\ models (black), which have  spectra similar to LBVs (see also \citealt{gme13}). The 2002 July observations of the LBV AG Car (green) from \citet{ghd09} are also show. Note that at that time, AG Car did {\it not} show a WN11h spectral type, since it had $\teff\simeq16400~$K \citep{ghd09}. {\it (b)}: The 28~\msun model (black) compared to the observations of AG Car in 1989 June (from \citealt{stahl01}, when it had a WN11h spectral type). {\it (c)}: The 32~\msun\ (orange), 40~\msun\ (green), 60~\msun\ (blue), 85~\msun\ (red), and 120~\msun\ (black) models, which have WO1--3 spectral type. The optical spectrum of the Galactic WO 3 star WR 93b (cyan) is also displayed. }}
\end{figure*}

Figures \ref{figspec1} and \ref{figspec2} show the synthetic optical spectra of SN progenitors with $\mini=20~\msun$ to 120 \msun. We display representative spectral regions that allow a broad illustration of the spectral morphology and spectral type determination. The ultraviolet to infrared spectra are available in the online version. To classify the synthetic spectra, we used the criteria from \citet{crowther98} for WO and WC stars, and \citet{ssm96} and \citet{crowther95a} for WN stars. Models that have spectrum similar to observed bona-fide LBVs, such as AG Car, P Cygni, and HR Car, have their spectral type listed as LBVs. While we recognize that formally there is no ``LBV'' spectral classification, we opted to use this classification since there is no objective spectral classification criteria of stars with dense winds that have $8000 \K \lesssim \teff \lesssim 25000 \K$. The spectra of these stars have  been commonly referred to in the literature as ``P Cygni-type'', ``iron'', and ``slash'' stars (see, e.g., \citealt{wf2000,clark12a}). The results of our spectral classification of the SN progenitors are summarized in the last column of Table \ref{model log}.

The pre-SN optical spectrum of the non-rotating model with \mini=$25~\msun$ is displayed in Fig. \ref{figspec1}a. The spectrum is characterized by strong emission of  \ion{H} {i} lines with shallow P-Cyg absorption profiles. Moderate emission of \ion{He}{i},  \ion{N}{ii} , and  \ion{Si}{iv} lines are also seen, while \ion{N}{iii} lines are weak or absent. Following the criteria from \citet{crowther95a}, the presence of H, the  \ion{N}{ii} lines being stronger than those of \ion{N}{iii}, and the presence of  \ion{He}{ii} $\lambda4686$ indicate  a WN11h classification. Indeed, the non-rotating $25~\msun$ model spectrum is similar to that of the LBV AG Carinae at visual minimum (1989 June), when it had a WN11h spectral type \citep{sc94,leitherer94,stahl01,ghd09}. The main difference is that the model spectrum has broader lines, since its $\vinf=694~\kms$ is about twice that of AG Car in 1989 June (300~\kms, \citealt{ghd09}). Another marked difference is the ratio between H and  \ion{He}{i} lines, which is smaller in the model spectrum compared to the observations of AG Car. This is due to a combination of higher He abundance and higher $\teff$ in the model spectrum compared to those of AG Car in 1989 June. We find that the non-rotating 25~\msun\ model has a peculiar appearance, looking something in between an LBV and WN11h star, and classify it as WN11h/LBV. Nevertheless, we note that all known WN11h stars are LBVs (or candidates) at their visual minimum, such as AG Car, Hen 3-519, and HDE 316285, so the spectral type WN11h seems to be exclusively linked to LBVs.

The synthetic spectra of the non-rotating models with $32~\msun < \mini < 40~\msun$ are shown in Fig. \ref{figspec1}b. Our models indicate that the non-rotating $32~\msun$ and 40~\msun\ models have at their final stage an optical spectrum that shows strong \ion{He}{ii} \lam4686, \ion{N}{iii} \lam4640, and \ion{N}{iv} \lam4057. Numerous \ion{He}{i} lines are also present with strong P-Cygni profile, such as \ion{He}{i} \lam5876 and \ion{He}{i} \lam6678. These features are characteristic of  a WN star of late spectral type (WNL). The He ionization structure indicates a WN 7 spectral type, but close to the transition to WN 8.  In our models, the continuum-to-peak ratio of \ion{He}{i} \lam5411/ \ion{He}{i} \lam5876$\simeq0.71$, while the boundary between WN7 and WN8 spectral types lies at  \ion{He}{i} \lam5411/ \ion{He}{i} \lam5876=0.65 \citep{ssm96}. The N ionization structure is broadly supportive of a WN7-8 classification. Our synthetic spectra have line widths not extremely broad, and are indicative of a WN7-8o subtype. For comparison, we present the observed spectrum of a typical Galactic WN7 (WR120) and WN8 star (WR 123), obtained from the atlas of \citet{hamann95}.

Figure \ref{figspec1}c displays the SN progenitor spectra from non-rotating models with $50~\msun < \mini < 120~\msun$. Their optical spectra are characterized by broad emission lines of \ion{O}{vi} \lam3811 and \ion{C}{iv} \lam5808. {\it This implies that they have WO spectral types at the pre-SN stage}, confirming previous indications that even at solar metallicity, massive stars can have their final stage as WO stars \citep{sander12,yoon12}. Here, we proceed one step further by computing the synthetic spectra, which allows us to derive spectral types and estimate the mass range where WO are SN progenitors. We obtain spectral types WO 1--3, depending on whether primary (ratio of EW of \ion{O}{vi} $\lambda$3818/\ion{O}{v}  $\lambda$5590) or secondary classification criteria (\ion{O}{vi} $\lambda$3818/\ion{C}{iv}  $\lambda$5808) are employed (see \citealt{crowther98}). Nevertheless, it is clear that the SN progenitors with $50~\msun < \mini < 120~\msun$ should be the hottest massive stars known.

We note that the synthetic spectrum of the non-rotating 120~\msun\ model has emission line strengths comparable to those of observed Galactic WO stars \citep{drew04,sander12}. Since the model has not been fiddled to fit the observations, several differences are readily noticeable between the observed and synthetic spectrum. This concerns mostly the \ion{C}{IV} $\lambda$5810 line, which is weaker in the models than in the observations. This is likely caused by the  hotter $\teff$ of the models, which shift the C ionization structure towards more ionized ions, making  \ion{C}{IV} $\lambda$5810 weaker.  

\clearpage

\clearpage

\clearpage

The synthetic spectra of the other WO models have much weaker emission lines compared to the observations. This striking result could be interpreted as theoretical evidence for a yet undiscovered population of weak-lined WO stars in the Galaxy. Alternatively, it could be linked to the \mdot\ prescription adopted by the Geneva code and used as input in CMFGEN to compute the synthetic spectra. Nevertheless, we find that models with $\mdot$ increased by a factor of 2 (or \vinf\ decreased by a similar amount) would still produce a WO spectrum, but with line strengths similar to the observations. To illustrate this effect, we show in Fig. \ref{figspec1}c the spectrum of  the 60~\msun\ model with \mdot\ increased by a factor of two (orange), while the other parameters are kept fixed. Without further evidence for a larger (or lower) $\mdot$ at the WO stage and for consistency with the stellar evolution models, we chose to keep the values of $\mdot$ used in the Geneva code and quoted in Table \ref{model log}.

The optical spectra of the  $20~\msun$ and 25~\msun\ rotating models are presented in Fig. \ref{figspec2}a. They have been discussed in \citet{gme13}, where we have shown that these stars, at the pre-SN stage, look much alike LBVs such as AG Car \citep{ghd09,ghd11}, HR Car \citep{gdh09}, P Cygni \citep{najarro97,najarro01}, and HDE 316285 \citep{hillier98}, among others. Their optical spectrum is dominated by strong lines of H, \ion{He}{i}, \ion{N}{ii}, and \ion{Fe}{ii}, often with P-Cygni profiles.

Figure \ref{figspec2}b shows the optical spectrum of the pre-SN rotating 28~\msun\ model, which has strong  \ion{He}{i} lines with P-Cygni profiles. Moderate  \ion{He}{ii} $\lambda4686$ and \ion{Si}{iv} emissions are  also present, while  \ion{N}{ii} and  \ion{N}{iii} lines have similar strength. This seems to be a transition spectral morphology between a WN10 and WN11, without the suffix 'h' since H lines are clearly not present. 

In Fig. \ref{figspec2}c we display the optical spectra of rotating models with $32~\msun < \mini < 120~\msun$ at the pre-SN stage. They show broad emission lines of \ion{O}{vi} \lam3811 and \ion{C}{iv} \lam5808, which also implies  a WO1--3 classification. The spectral morphology is similar to the non-rotating WO models. The rotating $32~\msun$ model has a strong-lined spectrum, which resembles the observed Galactic WO stars. The 40, 60, 85 and 120~\msun\ models have much weaker emission lines. Again,  this may be a real effect, which would make challenging to detect the more massive rotating models owing to their weak lines. Alternatively, $\mdot$ could be underestimated in the models.
 
While we did not compute spectra for the models that end up as RSGs, both non-rotating and rotating models become hotter for larger $\mini$. The relationship between spectral types and $\teff$ of RSGs has been the subject of debate recently. Using the most recent calibration for Galactic RSGs from \citet{levesque05}, we find that the SN progenitors from non-rotating stars have spectral types that range from M5 I (9 \msun\ model) to M1 I (20 \msun), while the spectral type of the rotating models lie between  M4.5 I (9 \msun\ model) to M2.5 I (15 \msun). However, the recent results from \citet{davies13} casts serious doubts on the validity of the $\teff$ -- spectral type calibration of RSGs, both for the LMC/SMC and for the Galaxy (B. Davies et al., private communication). Therefore, we quote the spectral types of RSGs as ``K--M I".

As described in Sect. \ref{teffhr}, the rotating 18~\msun\ model is a yellow luminous star at the pre-SN stage, with $\teff=5382~\K$ and $\lstar=1.5\times10^5~\lsun$.  We shall comment on its spectral type, even though we are not able to compute a synthetic spectrum for this model since its $\teff$ is too low for a CMFGEN analysis. Stars finishing their lives as yellow luminous stars have been customarily classified as yellow supergiants (YSG); see e.g. \citet{eliasrosa09,eliasrosa10,fraser10,maund11,georgy12}. However, we suggest here that some of these stars are actually yellow hypergiants (YHG). YSGs and YHGs share a similar region of the HR diagram, having similar $\teff $ and $\lstar$ \citep{dejager98}. The main difference is that, contrary to YSGs, YHGs have H$\alpha$ emission and broader absorption lines. This points out to the presence of a strong mass outflow and an unstable, extended atmosphere close to the Eddington limit in YHGs. Quoted values of \mdot\ of YHGs range from  0.2 to $2 \times10^{-5}~\msunyr$ during quiescence \citep{dejager98}, as in the case of $\rho$ Cassiopeiae \citep{lobel03}, IRC +10420 \citep{driebe09}, and W4 in Westerlund 1 \citep{dougherty10}. Here we find that the 18~\msun\ rotating model has a quite strong wind at the pre-SN phase with $\mdot=1.7\times10^{-5}~\msunyr$. This is well in line with the values quoted above for YHGs, which makes it very likely that an extended atmosphere is present. In addition, the model is also extremely bright (Sect. \ref{absmag}). Therefore, we propose a YHG classification with spectral type G1 Ia$^+$ following the $\teff$ $vs.$ spectral type calibration of \citet{kov07}.

\subsection{SN types associated with the progenitors}
Now that we have determined the spectral type of core-collapse SN progenitors in the initial mass range of 9--120~\msun\,  we would like to know to which kind of SN they lead to. Ideally, one would need to use the progenitor structure from our stellar structure models as input for hydrodynamical and radiative transfer simulations of the SN (if one indeed occurs). This is well beyond the scope of this paper, and here we use the chemical structure of the progenitor (in particular H and He) as a diagnostic for the SN type. As has been discussed by others (e.\,g., \citealt{heger03}, \citealt{georgy09},  \citealt{dessart11b}, \citealt{eldridge13}), this is a challenging task, since it is not fully understood how much H exists in the ejecta of SN IIP, IIL, and IIb, and the He content in SN Ib and Ic. Here, we assume the initial mass ranges for the different kind of SNe determined by \citet{georgy12a}, using the chemical abundance criteria as noted in Sect. \ref{ccsnfrac} of the present paper. The analysis that follows below has also the caveat that we assume that all core collapse events lead to a SN, even if a black hole is formed. In addition, as shown by \citet{georgy12}, mass loss during the RSG phase has a significant impact in the position of the star in the HR diagram at the pre-SN stage. Here, we analyze only models with the standard mass-loss rate during the RSG phase as described in \citet{ekstrom12}.

\begin{figure}
\center
\resizebox{0.995\hsize}{!}{\includegraphics{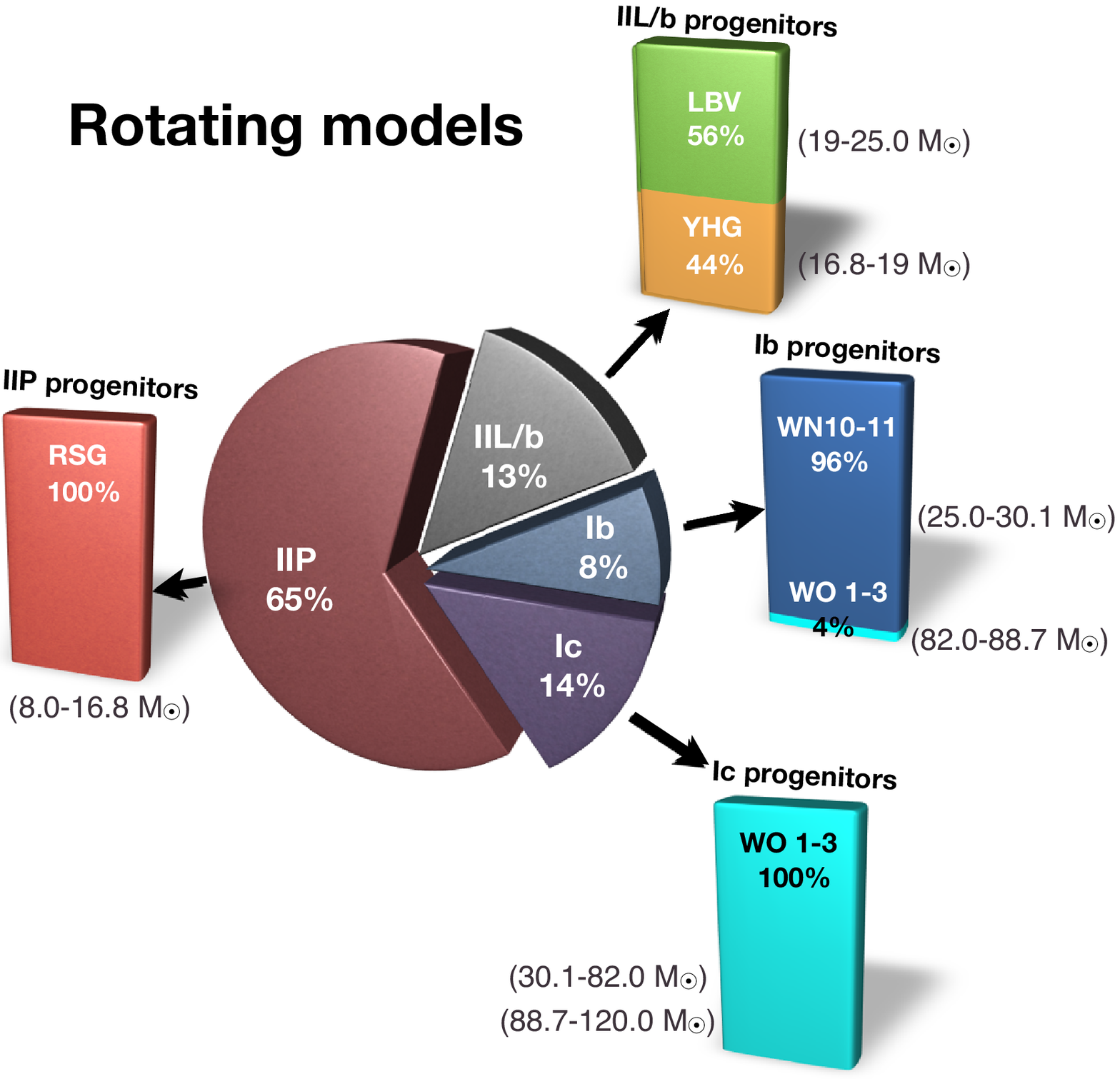}}\\
\resizebox{0.995\hsize}{!}{\includegraphics{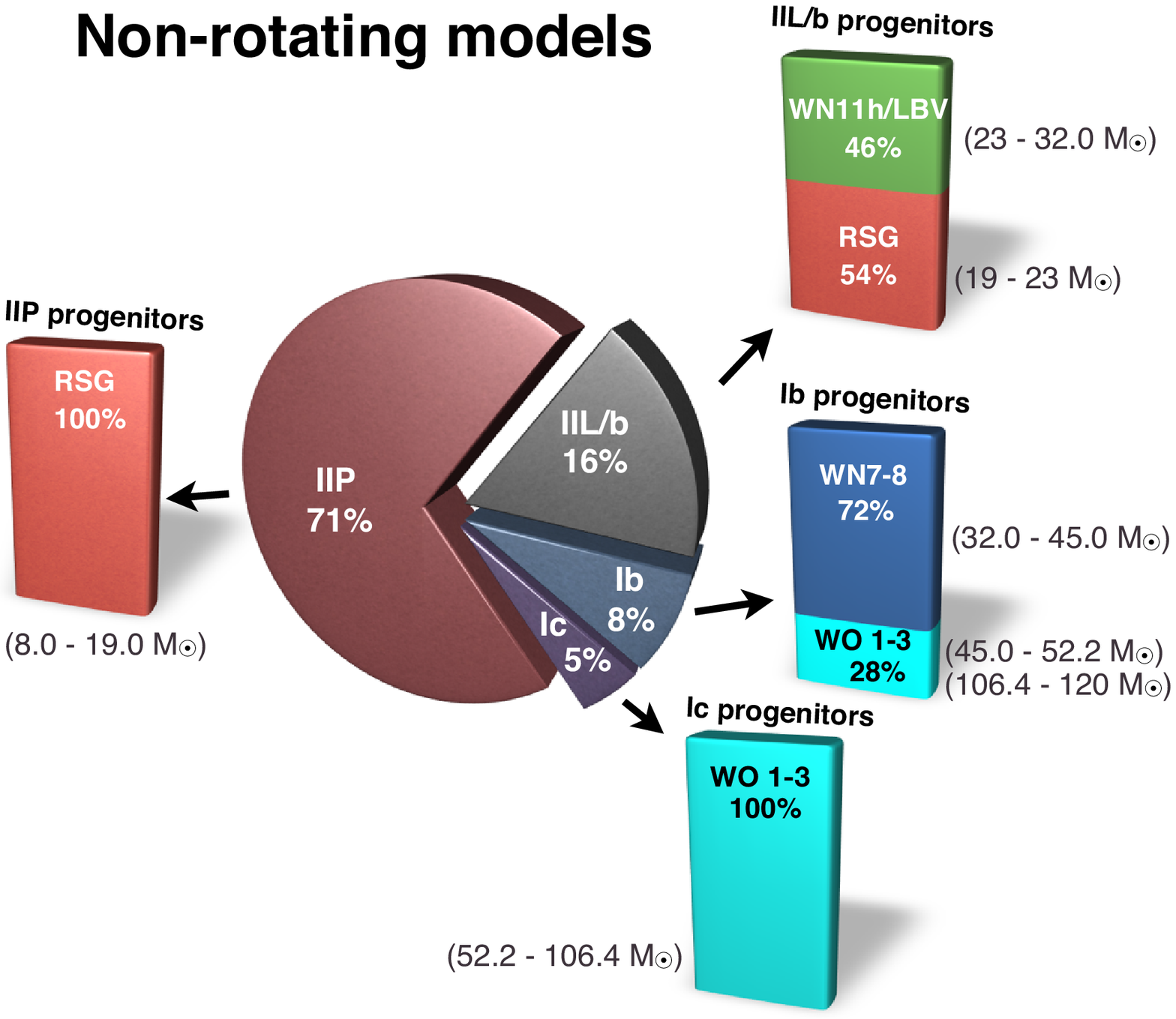}}
\caption{\label{diagfrac} Diagram illustrating, for different  core-collapse SN types, their relative rates and the types of progenitors and their respective frequencies. Initial mass ranges (indicated in parenthesis) and SN types are based on the criteria outlined in \citet{georgy12a}, assuming that the minimum amount of He in the ejecta to produce a SN Ib is 0.6~\msun. The upper and lower panels refer to rotating and non-rotating models, respectively.}
\end{figure}

Our results are summarized in Fig. \ref{diagfrac}, which presents the different channels that link the spectral type of SN progenitors to the core-collapse SN types. The mass ranges of progenitors leading to the different SN types are also shown in parenthesis. In some cases, it is possible that progenitors with different spectral types lead to the same SN type. We then assume a Salpeter-type initial mass function to obtain the fraction of the SN progenitors that come from the different channels.  

We first discuss the results of the rotating models. We obtained that 100\% of the SN IIP progenitors are RSGs. For the SN IIL/b progenitors, we found that 44\% of them are YHGs while 56\% are LBVs. Our definition supposes that SN progenitors with $4500 < \teff < 7000$ K are YHGs and, since there is an overlap in spectral type between LBVs and YHGs, we stress that the numbers above are indicative only. The progenitors of SN Ib are in the vast majority WN 10--11 stars, with only 4\% coming from WO stars.   Our models indicate that the SN Ic progenitors are all WO stars. 

Interestingly, the 25~\msun\ model, which has an LBV spectrum, is at the transition where stars lose all the H envelope. It represents likely a transition between an SN IIL/b and SN Ib. It thus seems that there is a parameter space where a small fraction of stars with spectral type similar to LBVs could explode as SN Ib. This is of interest to explaining the progenitor of SN 2006jc, which had an outburst a few years before core collapse was a type Ibn \citep{pastorello07}.

For the non-rotating models, the nature of the SN IIP progenitors is unchanged, with all of them being RSGs. However, we found that SN IIL/b progenitors could come from RSGs (54\%) or WN11h/LBV stars (46\%). The nature of the SN Ib progenitors is  changed, but the majority of the progenitors are still late WN (WN 7--8) stars, and 28\% are WO stars. The nature of the SN Ic progenitors is unchanged as compared to rotating models, with 100\% of the progenitors being WO stars. These stars present a WO-type spectrum only during a very short period preceding their core collapse, therefore this finding is not in contradiction with the fact that WO stars are very rare stars (see Sect. \ref{wosn}).

\subsection{Linking the interior properties of SN progenitors and their spectra}

\begin{figure*}
\center
\resizebox{0.995\hsize}{!}{\includegraphics{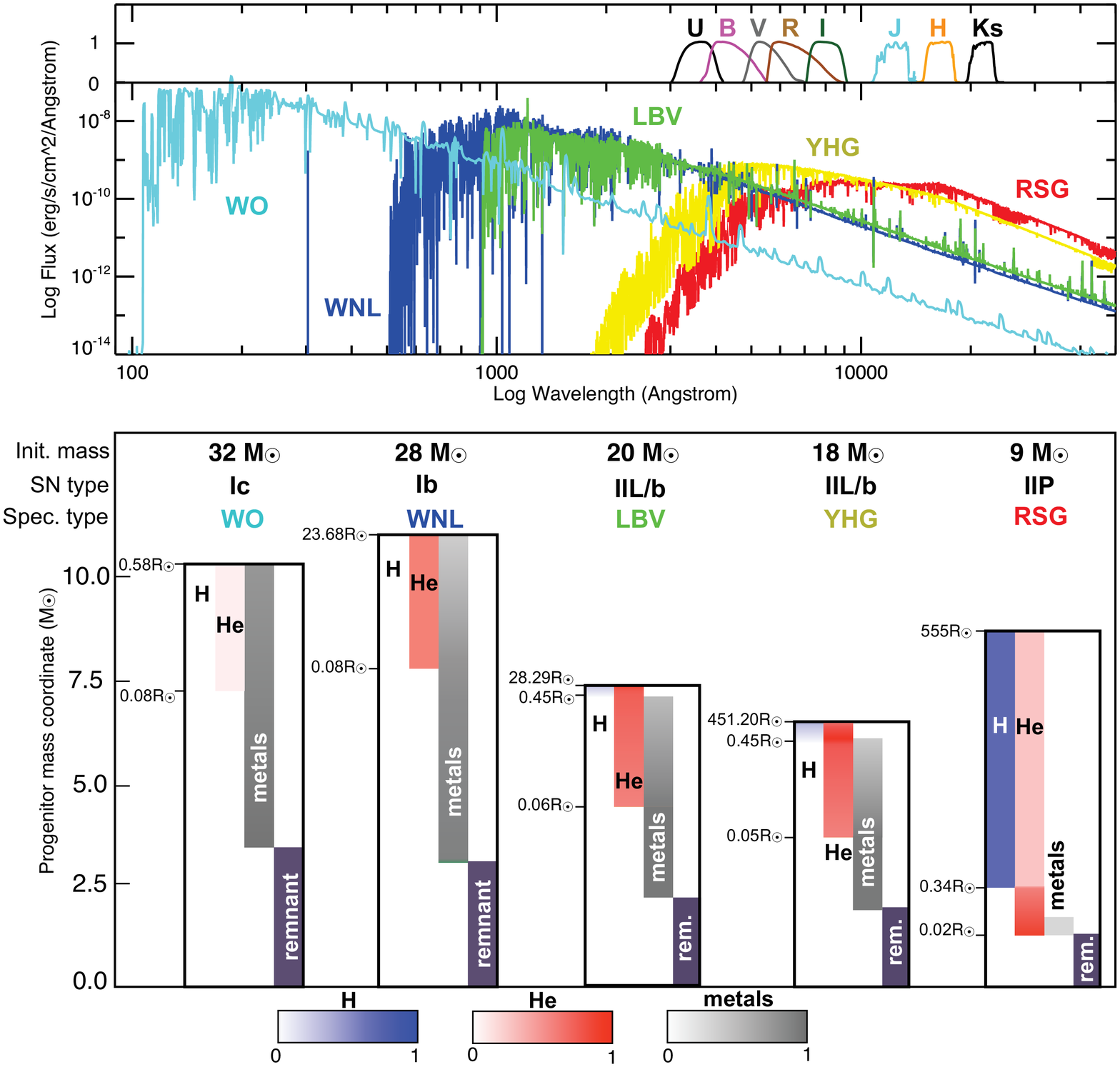}}
\caption{\label{snprogsed} {\it Top:} Spectral energy distribution of selected models, showing a WO star with $\teff \simeq154000~\K$ (our 32~\msun\ rotating mode; cyan), a WNL star of spectral type WN10--11 with $\teff \simeq26800~\K$ (28~\msun\ rotating model; blue), and an LBV with $\teff\simeq20000~\K$ (20~\msun\ rotating model; green ). We stress that we do not compute model spectra of RSG and YHGs in this paper. To illustrate the SED of these objects, we also overplot MARCS model spectra of a RSG with $\teff=3600~\K$ and luminosity scaled to $\lstar=1.2\times10^5~\lsun$ (red), and of a YHG with $\teff=5250~\K$ and luminosity scaled to $\lstar=1.5\times10^5~\lsun$ (yellow). All fluxes have been arbitrarily scaled to a distance of 1~kpc. The upper inset shows the normalized bandpasses of the $UBVRIJHK_S$ filters. {\it Bottom:} Schematic illustration of the interior structure of the SN progenitors for which spectra are shown in the upper panel. We show the Lagrangian mass coordinate of the progenitor in the y axis and the extension of the layers  for different chemical elements (H, He, metals) and the baryonic remnant mass (computed as in \citealt{hirschi05}). The radius of selected shells is indicated on the left side of each sub panel, while the spectral type, SN type, and initial mass of the progenitors are indicated immediately above the interior structure. The chemical abundances (by mass) are color-coded  in blue (H), red (He), and grey (metals). }
\end{figure*}

Having determined the SN type associated with each progenitor spectral type, let us investigate how the interior structure of SN progenitors is related to their spectra. Figure \ref{snprogsed} presents in a schematic way both the spectral appearances of five selected supernova progenitors (upper panel), together with information about their internal structures (lower panel). 

Looking at the structure, we readily note that despite their different surface properties and spectra,  all progenitor models have an He-free core extending over a radius of a few hundredths of~\rsun. The outer  layers of the He-free cores are made up of  C and O produced by the He-burning reactions. For this reason, this region is called the carbon-oxygen core. Of course the very central part of this zone is no long composed of C and O, since the advanced stages have further processed these elements into heavier species. 

We see that the carbon-oxygen cores vary in size from two to nine times the radius of the Earth when going from the progenitor that is a RSG to the one that is a WO. At this stage, this part of the star is already sustained at least in part  by electron degenerate pressure (remind that a white dwarf has typically the mass of the Sun locked into a sphere having the radius of the Earth).  These cores have masses well above the Chandrasekhar mass and are on the verge of collapsing under the effect of their own gravity. 

Let us now consider the three models associated to the LBV, YHG and RSG progenitors.  Interestingly, the radii of the H-free cores (i.e. He cores) are very similar in the three models and are equal to a few tenths of solar radius. As a rule of thumb, we see that the H-free core is roughly one order of magnitude larger than the He-free core. 

In contrast with the cores, the envelopes of the these three progenitors differ greatly. While in the RSG progenitor, the envelope contains more than 6~\msun\ and is quite H-rich, the one in the YHG only comprises a fraction of a solar mass and is much less H rich. The trend continues with a decrease of the mass of the envelope and its H content when one passes from the YHG to the LBV progenitor. 

Strikingly, the extensions of the envelopes of the RSG and YHG are not so different and are equal to a few hundreds of solar radii, while as we have just noticed their masses and H contents are very different. This also illustrates the fact that removing a large part of the H-envelope and reducing significantly its H content is not sufficient to make the star evolve back to the blue part of the HR diagram. For this to occur, strong mass loss is needed, so the H content of the envelope is reduced to a few hundreths of \msun. This is well confirmed by many previous works \citep{giannone67,vanbeveren98a,salasnich99,yooncantiello10,georgy12,meynet13}. 

As said above, if the reduction of the H-rich envelope continues, the star will evolve in the blue becoming an LBV or a WR star. We see that our LBV progenitor has a structure not so different from the YHG except for the fact that the LBV has a reduced H-rich envelope. 

The LBV and WNL progenitors have similar structures and similar radii, while a drastic change in radius occurs when one passes from the WNL to the WO progenitor. Note that H content of the outer layers seems to be one of the key factors leading to these various supernovae progenitors.  The sequence of models from right to left corresponds to a decrease of H in the envelope. It seems that the passage from red to blue (and from blue to very blue) does not occur gradually but rather in a very sharp manner when some limit in the H surface content is reached.

\section{Spectroscopic vs. chemical abundance classifications of massive stars}
\label{specchem}
\begin{table*}
\begin{minipage}{\textwidth}
\footnotesize
\caption{Surface chemical composition and comparison of spectral types of core-collapse SN progenitor with previous studies.}
\label{comp}
\centering
\vspace{0.1cm}
\begin{tabular}{l c c c c c c c c}
\hline\hline
M$_\mathrm{ini}$&   $X_\mathrm{surf}$ & $Y_\mathrm{surf}$ & $C_\mathrm{surf}$ & $N_\mathrm{surf}$ & $O_\mathrm{surf}$ & Previous Sp. Type\tablefootmark{a} & New Sp. Type& SN type\tablefootmark{a} \\ 
(\msun)		& 		(mass)             & 		(mass)        & 		(mass)        & 		(mass)        & 		(mass)        & (Chem. Abund.) &  &\\
\hline 
 \hline
 \multicolumn{9}{c}{non-rotating models }\\
 \hline 
           9 &  0.68  & 0.31  & 1.6$\times10^{-3}$ & 3.0$\times10^{-3}$ & 4.4$\times10^{-3}$ & RSG & $K-M~I  $ & IIP \\
          12 &   0.68  & 0.31  & 1.3$\times10^{-3}$ & 2.5$\times10^{-3}$ & 4.8$\times10^{-3}$ & RSG & $K-M~I $   &  IIP  \\
          15 &  0.65  & 0.34  & 1.1$\times10^{-3}$ & 3.2$\times10^{-3}$ & 4.4$\times10^{-3}$ & RSG &  $K-M~I$    &  IIP \\
          20 &  0.48  & 0.51  & 8.4$\times10^{-5}$ & 6.9$\times10^{-3}$ & 1.6$\times10^{-3}$ & RSG & $K-M~I$    &  IIL/b  \\
          23 &   0.31 & 0.67  & 6.4$\times10^{-5}$ & 8.1$\times10^{-3}$ & 1.5$\times10^{-4}$ & RSG &  $K-M~I$     & IIL/b\\                                                                                                                        
          25 & 0.16& 0.83    &6.8$\times10^{-5}$ & 8.2$\times10^{-3}$ &1.1$\times10^{-4}$ & WNL & WN11h/LBV &    IIL/b    \\
          32 & 0.00& 0.99    &1.4$\times10^{-4}$ & 8.1$\times10^{-3}$ &1.0$\times10^{-4}$ & WNE & WN 7--8o              &   Ib\\
          40 & 0.00& 0.99    &9.5$\times10^{-5}$ & 8.2$\times10^{-3}$ &9.2$\times10^{-5}$ & WNE &  WN 7--8o            &  Ib \\
          50 & 0.00 & 0.19   &0.46 & 1.5$\times10^{-5}$ & 0.33 & WC & WO 1--3 					                     & Ib \\
          60 & 0.00& 0.27    &0.51 & 0.00 &0.20 & WC & WO 1--3  													&  Ic \\
          85 & 0.00& 0.25   &0.49 & 0.00 &0.25 & WC & WO 1--3 														 &  Ic \\
         120 & 0.00& 0.24   &0.46 & 0.00 &0.29 & WC &WO 1--3														 &  Ib\\       
  \hline
 \multicolumn{9}{c}{rotating models, initial $\vrot/\vcrit=0.4$ }\\
 \hline 
           9 &   0.64& 0.35  &6.2$\times10^{-4}$ & 3.8$\times10^{-3}$ & 4.4$\times10^{-3}$ & RSG &$ K-M~I$   &  IIP      \\
          12 & 0.64& 0.35  &6.2$\times10^{-4}$ & 3.9$\times10^{-3}$ & 4.2$\times10^{-3}$ & RSG & $K-M~I $      &  IIP        \\
          15 & 0.56& 0.43  &4.9$\times10^{-4}$ & 4.7$\times10^{-3}$ & 3.5$\times10^{-3}$ & RSG & $K-M~I$        &  IIP  \\
          16.5 & 0.55& 0.44  &4.8$\times10^{-4}$ & 4.8$\times10^{-3}$ & 3.4$\times10^{-3}$ & RSG &$K-M~I$     &  IIP  \\
          18 & 0.38& 0.61  &4.9$\times10^{-5}$ & 7.0$\times10^{-3}$ & 1.4$\times10^{-3}$ & RSG &$G1 Ia^+$       &  IIL/b  \\
          20 & 0.24& 0.75  &6.2$\times10^{-5}$ & 7.8$\times10^{-3}$ & 4.9$\times10^{-4}$ & WNL & LBV  			&  IIL/b   \\
          25 & 0.03& 0.93  &1.6$\times10^{-4}$ & 1.6$\times10^{-3}$ & 6.5$\times10^{-4}$ & WNL &  LBV				     	 &  Ib    \\
          28 & 0.00& 0.97  &9.3$\times10^{-3}$ & 1.0$\times10^{-2}$ & 1.2$\times10^{-3}$ & WNL		& WN 10--11             &  Ib    \\
          32 &  0.00& 0.28  &0.47 & 2.2$\times10^{-3}$ &0.23 & WC & WO 1   								        &  Ic   \\
          40 &   0.00& 0.28  &0.50 & 4.0$\times10^{-5}$ &0.20 & WC & WO 1-2										  &  Ic   \\
          60 &  0.00& 0.28  &0.49 & 0.00 &0.21 & WC & WO 1-3													  &   Ic \\
          85 &  0.00& 0.26  &0.47 & 0.00 &0.25 & WC &WO 1-3  													  &  Ib \\
         120 & 0.00& 0.26  &0.49 & 0.00 &0.23 & WC & WO 1   													  &  Ic \\
 \hline
\end{tabular}
\tablefoot{\tablefoottext{a}{From \citet{georgy12a}.}}
\end{minipage}
\end{table*}

We computed here, for the first time, the fundamental parameters and observables of SN progenitors across a broad mass range, based on evolutionary calculations. Not surprisingly, having a spectrum allows a much more precise determination of the spectral type of the SN progenitors, using the same observational criteria employed for classifying the spectra of real stars. This allows a refined understanding of which kind of single stars will be progenitors of the different types of core-collapse SNe.

In previous studies, the properties of core-collapse SN progenitors, such as their spectral type, were estimated according to the surface chemical abundances, luminosity, and temperature \citep{georgy09,georgy12a}. These were accomplished using the relationship between spectral types and surface abundances from \citet{smithmaeder91} and \citet{meynet03}, as follows: 
\begin{itemize}
\item RSGs have $\log(T_\text{eff}/\text{K}) < 3.66$;
\item O stars have a surface hydrogen mass fraction $X_\text{H} > 0.3$ and $\log(T_\text{eff}/\text{K}) > 4.5$;
\item WR stars have $\log(T_\text{eff}/\text{K}) > 4.0$ and $X_\text{H} < 0.3$;
\item WNL stars are WR stars with $X_\text{H} > 10^{-5}$.
\item WNE stars have $X_\text{H} < 10^{-5}$ and a C surface abundance $X_\text{C}$ lower than the 0.1 times the N surface abundance $X_\text{N}$;
\item WNC stars have  $X_\text{H} < 10^{-5}$ and $\frac{X_\text{C}}{X_\text{N}}=0.1$ to 10;
\item WC stars have $X_\text{H} < 10^{-5}$, $X_\text{C} > {X_\text{N}}$, and  surface abundances (by number) such as $\frac{\text{C} + \text{O}}{\text{He}} < 1$;
\item WO stars have  $X_\text{H} < 10^{-5}$, $X_\text{C} > {X_\text{N}}$, and $\frac{\text{C} + \text{O}}{\text{He}} > 1$.
\end{itemize}

In Table \ref{comp} we show how our refined spectral type classifications compared to those from \citet{georgy12a}. The abundances at the surface, which were the main criterium used in previous evolutionary studies, are also shown. Important differences between our spectral type determination and previous ones can be readily noted, and we comment on them below.

Starting from models with the lowest initial masses,  we note that RSGs were not classified in spectroscopic subtypes in previous stellar evolution studies. For the 20 and 25 \msun\ rotating models, we have determined that a spectral type reminiscent of LBVs is found. However, we see that a spectral classification of a late-type WN star (WNL) was suggested before. This was because, in the absence of a spectrum, it was assumed that stars with $\teff > 20000$~K and H surface abundance ($X_\mathrm{surf}$) greater than $10^{-5}$ would be WNL stars \citep{meynet03,georgy12a}. In \citet{gme13} we showed that the 20 and 25 \msun\ rotating models actually resemble LBVs rather than WNL stars. The WNLs with the latest possible spectral type (WN11h) have $\teff\gtrsim25000$K, which is well above the $\teff\simeq20000$~K found for the 20 and 25~\msun\ rotating models at the pre-SN stage. The non-rotating 25~\msun\ model is a transition between an LBV and a WN11h spectral type.

For higher initial masses, we show here that a WN 7--8 spectral type is found for the 32 and 40 \msun\ non-rotating models, while \citet{georgy12a} suggested an early-type WN (WNE) classification. This difference stems from the assumption made in previous works that stars with $X_\mathrm{surf} <10^{-5}$ would be WNE stars. With the spectra computed here, we show that this assumption is not necessarily true, and that WNL stars can also have $X_\mathrm{surf} <10^{-5}$. This was also found in previous works that analyzed observed WNL stars, such as WR 123 \citep{hamann06}. The main parameter that determines whether the star is a WNE or WNL is $\teff$, which regulates the He and N ionization structures in the wind and ultimately sets the spectral type of the star \citep{hillier87a}.

For the rotating 32 --120~\msun\ and non-rotating 60--120~\msun\ models we obtain a WO spectral type. This is also in contrast with previous estimates, that have commonly associated a WC classification to the SN progenitors based on the surface abundances of C, O, and He. Earlier studies assumed that a WO subtype would only appear when (C+O)/He$ > 1.0$ (by number) at the surface \citep{smithmaeder91}, i.e., that a significant O enrichment would be responsible for the appearance of the WO subtype. Here we find that, once a minimal amount of O is present at the surface, the WO subtype arises mainly because of ionization effects. The extremely high $\teff$ of stars at their end is what produces the broad, strong \ion{O}{vi} $\lambda$3818 emission that characterizes the WO subtype. Therefore, we show here that stellar evolution models predict, even at solar metallicity, that WO stars are the end stage of stellar evolution for the most massive stars, at least up to 120~\msun. Also, we note that WO stars arise both from non-rotating and rotating stars, although the mass range at which they appear is tighter for non-rotating  (60--120~\msun)  than for rotating stars (32--120~\msun). However, since rotating stars are more luminous than non-rotating stars of the same mass, the luminosity range in which WO appears is similar for rotating and non-rotating models.

In a further work, we shall compute the output spectra for all the evolutionary stages from the Zero Age Main Sequence to core collapse. This will allow us to provide the evolutionary connections of the spectral types as they are predicted by theoretical models for single stars.

\section{\label{absmag} Absolute magnitudes, colors, and bolometric corrections of SN progenitors}

Fortunately, a significant amount of archival high spatial resolution imaging data exists. When a SN occurs, these archival data can be used to search for the SN progenitor and obtain its magnitudes, or at least to derive lower limits. Here, we present a theoretical database to aid the comparison between the endstage of the Geneva evolutionary tracks and observations of SN progenitors. We performed synthetic photometry of our SN progenitor models using the Chorizos code \citep{chorizos04}, adopting its built-in passband and zero point definitions that were obtained from \citet{cohen03,ma06,holberg06,ma07}. 

For a given filter $P$, the absolute magnitudes ($M_P$)  in the modified Vega magnitude system are
\begin{equation}
M_{P} =  -2.5\log_{10}\left(\frac{\int P(\lambda)F_{\lambda}(\lambda)\lambda\,d\lambda}
                         {\int P(\lambda)F_{\lambda,\mathrm{Vega}}(\lambda)\lambda\,d\lambda}\right)
                       + {\rm ZP}_{P},
\label{absmageq}
\end{equation}
where  $\lambda$ is the wavelength, $P(\lambda)$ is the sensitivity curve of the system, $F_{\lambda}$ is the model flux at 10 pc, $F_{\lambda,\mathrm{Vega}}$ is the flux of Vega scaled to a distance of 10pc, and ${\rm ZP}_{P}$ is the zero point.

The bolometric magnitudes (\mbol) are computed with the usual relationship, assuming that the solar \mbol\ is 4.74 mag: 
\begin{equation}
M_{bol} =   -2.5\log_{10} \lstar + 4.74,
\label{mboleq}
\end{equation}
and bolometric corrections in a given filter $P$ (BC$_P$) are then obtained using
\begin{equation}
\mathrm{BC}_{P} = \mbol - M_{P}.
\label{bceq}
\end{equation}

To illustrate the wavelength dependence of the absolute magnitudes and bolometric corrections of our SN progenitor models, we quote them in the following filters:  Johnson-Cousins $UBVRI$, {\it Hubble Space Telescope (HST)}/Wide Field Planetary Camera 2 (WFPC2) $F170W$, $F300W$, $F450W$, $F606W$, $F814W$, and 2MASS $J$, $H$, and $K_S$, as can be seen  
in tables \ref{absmagbc1} and \ref{absmagbc2}.

\begin{figure*}
\center
\resizebox{0.95\hsize}{!}{\includegraphics{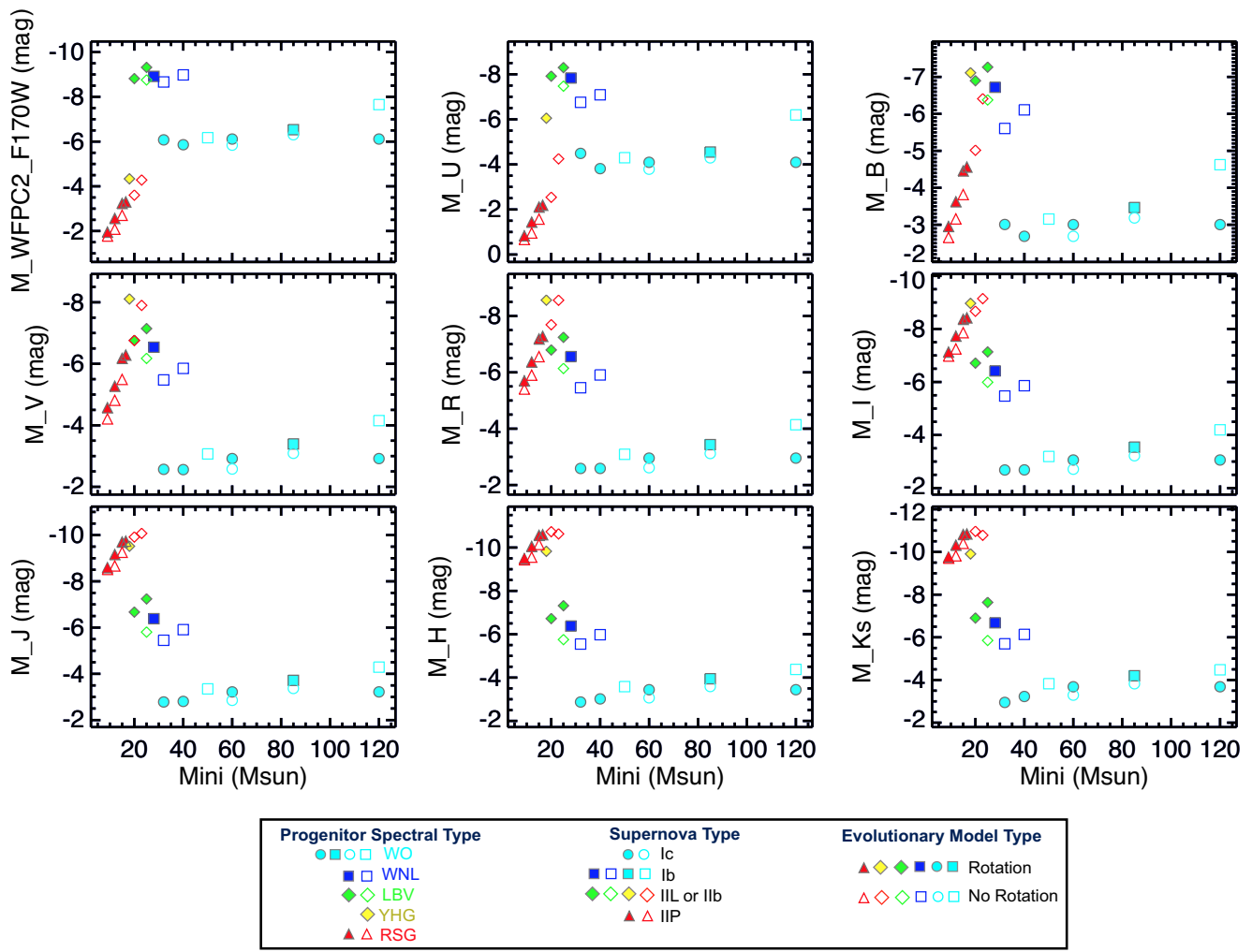}}
\caption{\label{figabsmagbc1} {Absolute magnitude of SN progenitor models as a function of initial mass, in the WFPC2/$F170W$ and $UBVRIJHK_S$ filters (from top to bottom, left to right).  We employ a three-dimensional scheme to label the symbols, according to the progenitor spectral type, SN type, and evolutionary model type (see lower panel). The progenitor spectral type is color coded, with cyan symbols corresponding to WOs, blue to WNL, green to LBVs, yellow to YHGs, and red to RSGs. The SN type is coded as the geometry of the symbol, with circles depicting progenitors of SN Ic, squares of SN Ib, diamonds of SN IIL/b, and triangles of SN IIP. The evolutionary model type is coded as open (without rotation) and filled (with rotation) symbols. }}
\end{figure*}

Figure \ref{snprogsed} (top panel) presents the spectral energy distribution of selected SN progenitor models, covering the range in $\teff$ explored in this work. The behavior of the absolute magnitudes and bolometric corrections as a function of initial mass is then regulated by how much flux from the star falls within the passband of a given filter. The shape of the spectral energy distribution depends on  the effective temperature, luminosity, and wind density. Therefore, not necessarily the most massive and luminous stars are the brighter ones in a given filter. This has also been discussed by \citet{eldridge07} in the context of  the comparison between RSGs and super Asymptotic Giant Branch stars. Figure \ref{figabsmagbc1} shows the absolute magnitudes in the WFPC2/$F170W$ and $UBVRIJHK_S$ filters as a function of initial mass. The symbols are coded according to the progenitor spectral type, ensuing SN type, and the evolutionary model type used (rotating or non-rotating).

Let us look at the absolute magnitudes and BCs of the SN progenitors and how they broadly vary according to the initial mass and wavelength. In general, we find that the RSGs (red symbols) are bright in the $RIJHK_S$ filters and faint in the $F170W$ and $UB$ filters. LBVs (green symbols), YHGs (yellow), and WNs (blue) are relatively bright in all filters, while WOs (cyan) are faint in all filters, with the exception of the $F170W$ filter.

We obtain that the brightest SN progenitors in the $F170W$  filter are the 20~\msun\ (LBV), 25~\msun\ (LBV), and 28~\msun\ (WN10--11) rotating models, and the 32 and 40~\msun\ non-rotating models (WN7--8o). They are followed in brightness by the 32--120~\msun rotating models and 50--120 non-rotating models, which are WOs. The RSGs are the faintest progenitors in these filters.

In the $U$ and $B$ bands, the brightest models are the 18, 20, 25, and 28~\msun\ rotating models, which have $\magu= -6.05, -7.92$, $-8.30$, and $-7.83$ mag, respectively (Fig. \ref{figabsmagbc1}). They are followed by the 25, 32, and 40~\msun\ non-rotating models (WN8, $\magu=-7.47$, $-6.76$, and $-7.09$, respectively). In the $U$-band, the 32 and 120~\msun\ rotating models (WOs)  are brighter than the rotating 9-15~\msun\ models (RSGs), while the situation is reversed in the redder filters. This is explained by the huge difference in $\teff$ between WOs ($\sim 145000-175000$~K) and RSGs ($\sim 3480-3740$~K). While the earlier emits the bulk of their flux in the far-UV, RSGs emit most of the flux in the near-IR (Fig. \ref{snprogsed}). The WO stars have similar absolute magnitudes, with the exception of the non-rotating 120~\msun\ model, which is significantly more luminous ($\sim0.8-1.3$ mag) than the other WO models.

In the $V$-band, we predict that the brightest SN progenitors should  be the 18~\msun\ rotating models (YHG) and 23~\msun\ non-rotating model (RSG). They are followed by the 20~\msun\ and 25~\msun\ rotating models, and by the non-rotating $20~\msun$ model. In the $R$- and $I$-bands, RSGs are the brightest progenitors, followed by YHGs, LBVs , WN8s, and WOs. In the near-infrared $JHK_S$, RSGs are even brighter than the other classes of progenitors.

\begin{figure*}
\center
\resizebox{0.95\hsize}{!}{\includegraphics{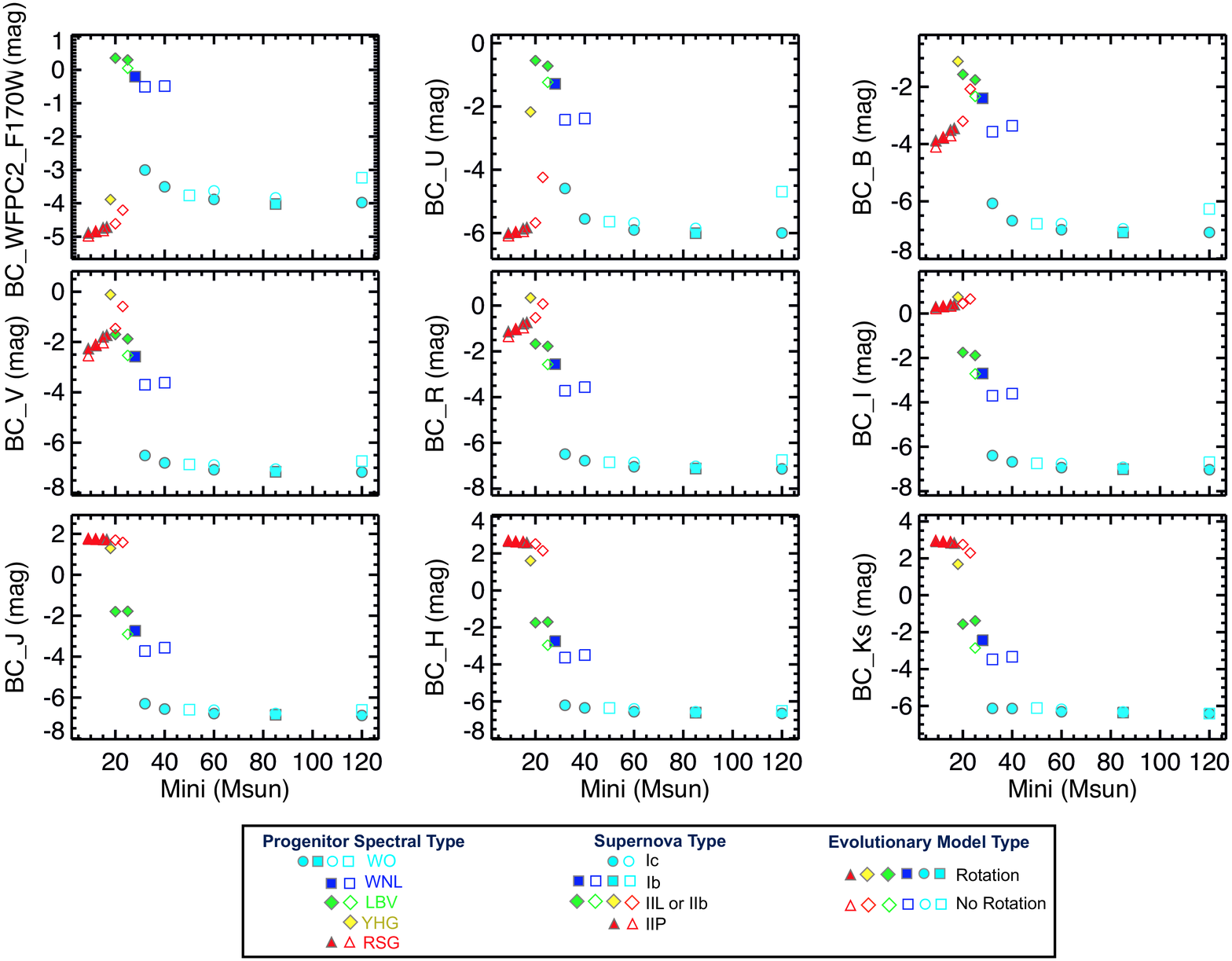}}
\caption{\label{figabsmagbc2} {Similar to Fig. \ref{figabsmagbc1}, but showing the bolometric corrections as a function of initial mass of the SN progenitors. }}
\end{figure*}

For progenitors in the RSG regime ($\mini=9-15 \msun$), the non-rotating models are systematically fainter than the rotating models at the same initial mass. The difference in magnitude can be up to 0.7 mag, as in the case of the 15 \msun\ model in the $V$-band. The 25~\msun\ non-rotating model is about one magnitude fainter than their rotating counterparts. Conversely, the non-rotating 32~\msun\ and 40~\msun\ models are significantly brighter in $UBVRI$ than the respective rotating models at the same initial mass, since the non-rotating models are significantly cooler (spectral type WN8) than the rotating models (spectral type WO1). Between 60~\msun\ and 85~\msun, the non-rotating models are slightly fainter than the rotating ones, while the non-rotating 120~\msun\ model is about 2 mag brighter than the rotating 120~\msun\ model. 

We  present in Fig. \ref{figabsmagbc2} the bolometric corrections in the different bands as a function of initial mass. These can be employed to obtain the bolometric luminosity in the cases where observations were performed in a single band, and one has indications of the nature of the progenitor. As one can see, the largest bolometric corrections (in modulus) are found for the WO stars, since they have most of their flux emitted outside the bandpass of the $UBVRI$ filters.

\begin{figure*}
\center
\resizebox{0.31\hsize}{!}{\includegraphics{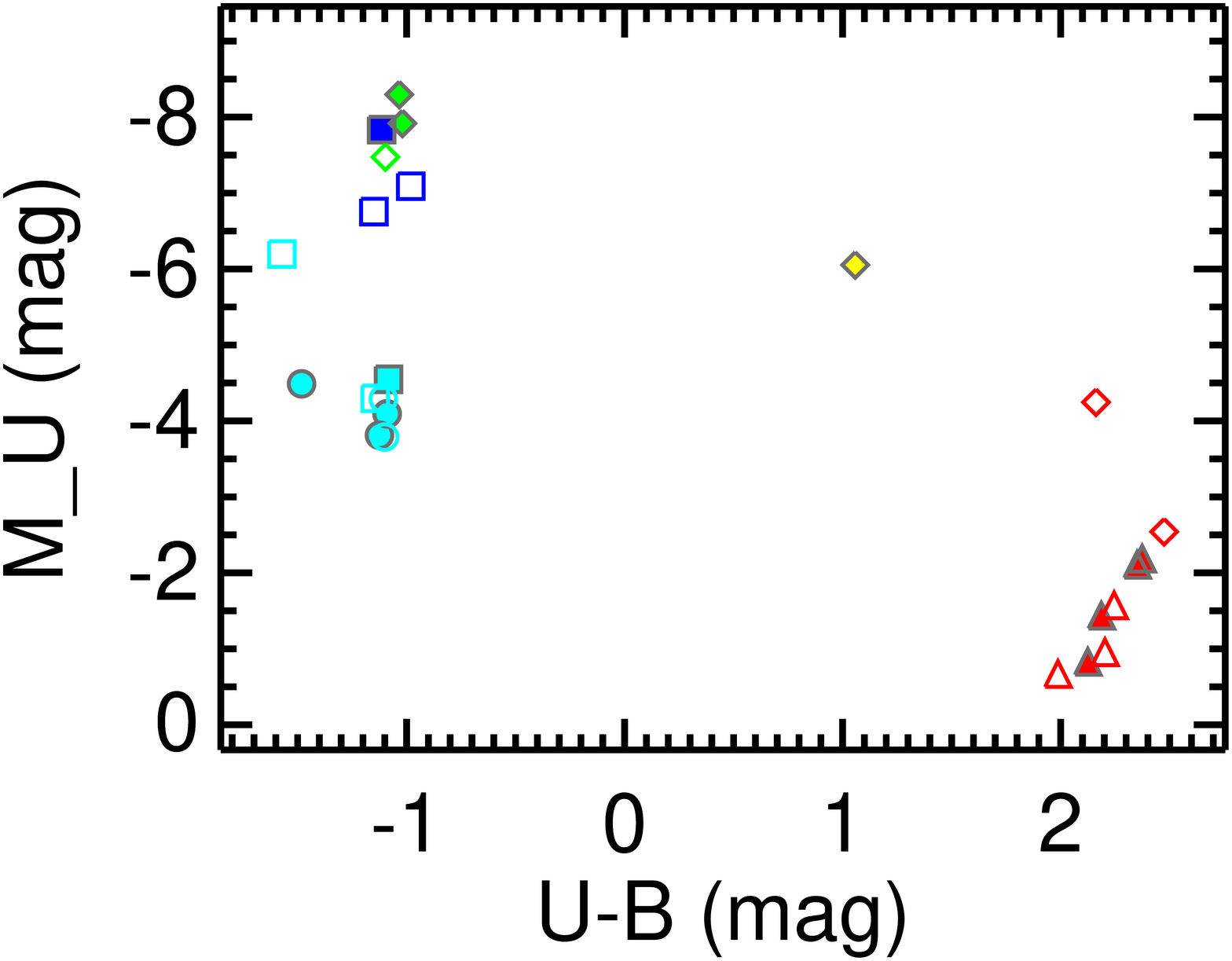}}
\resizebox{0.31\hsize}{!}{\includegraphics{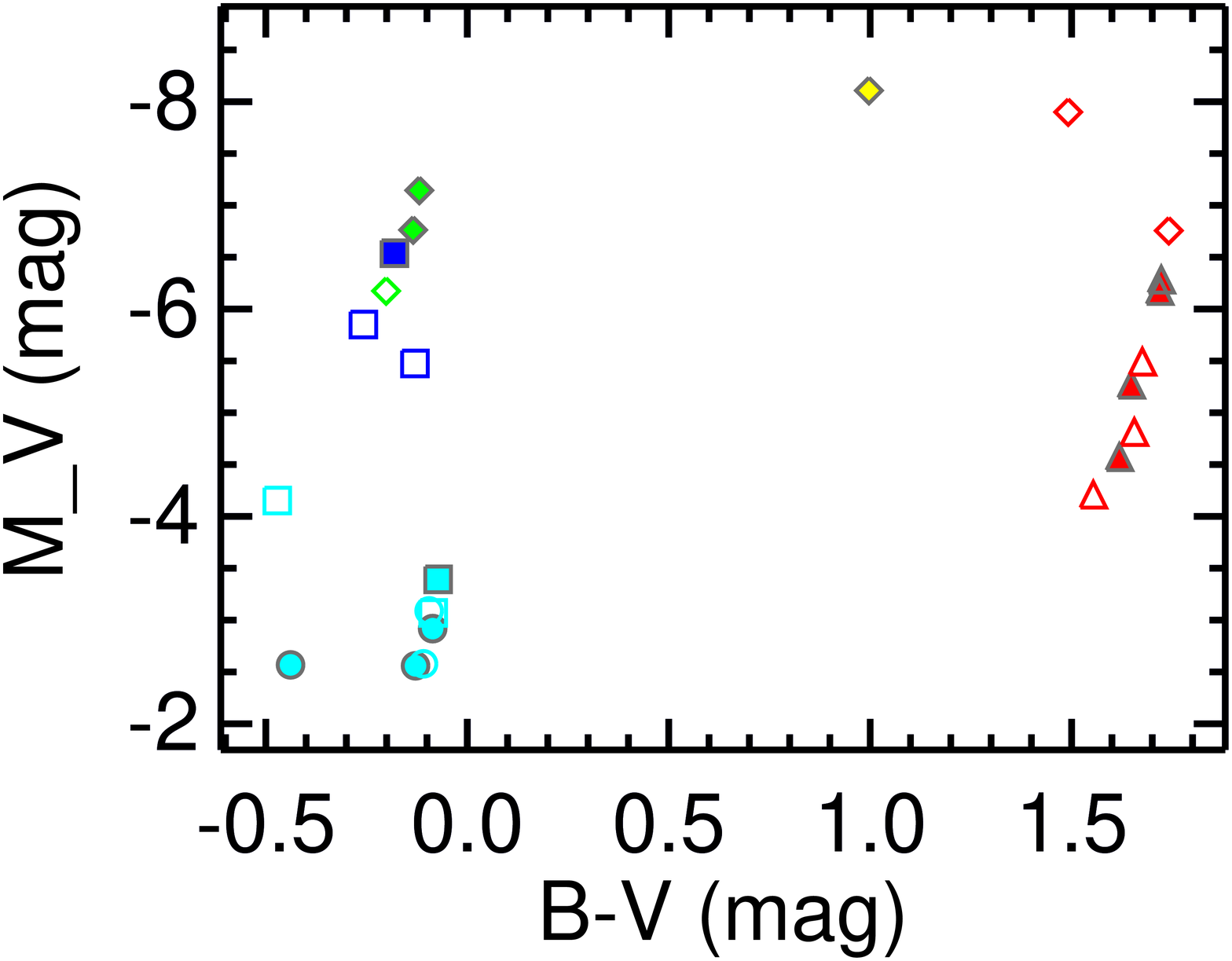}}
\resizebox{0.31\hsize}{!}{\includegraphics{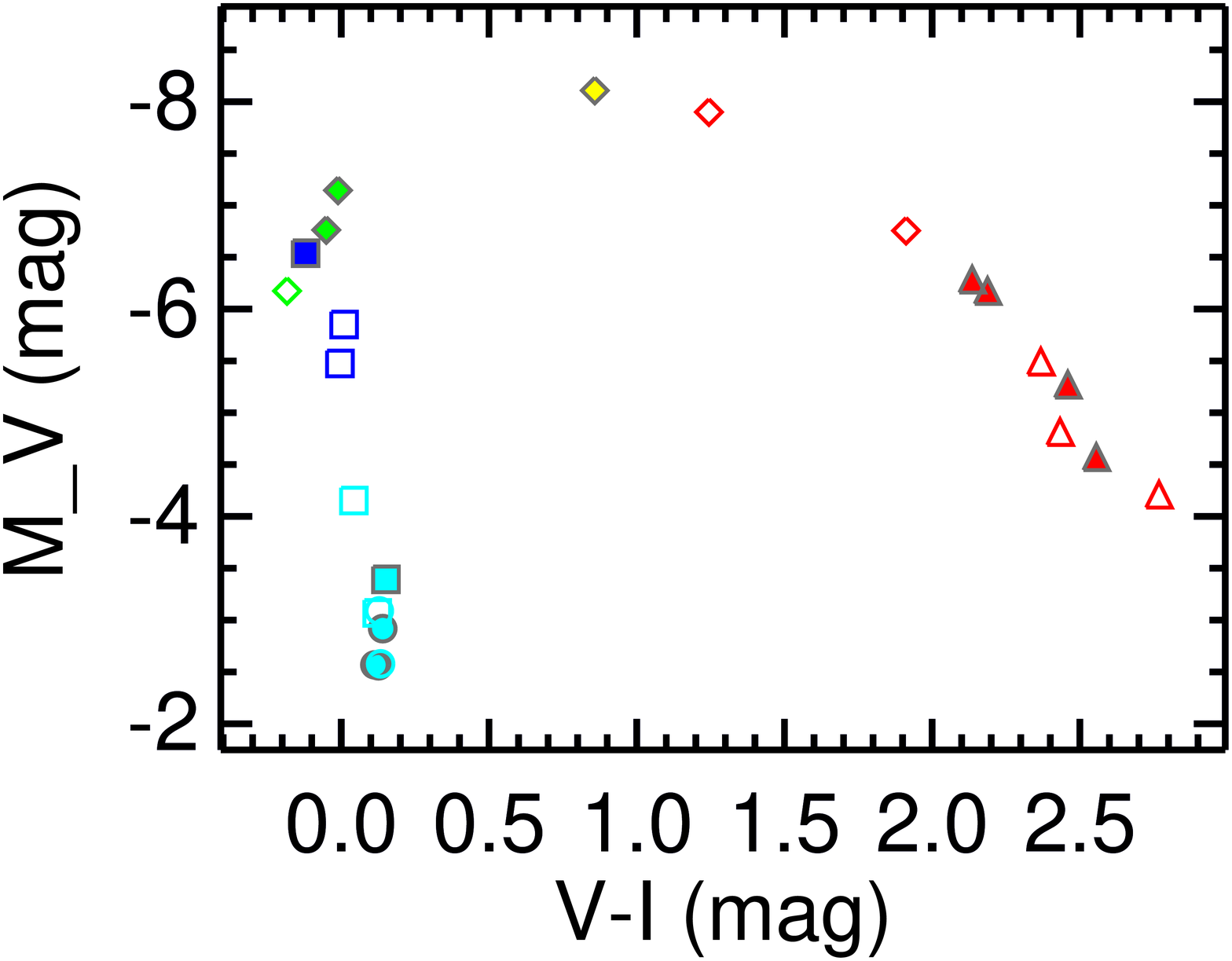}}
\caption{\label{figcolor} { From left to right, color-magnitude diagrams  $M_U~vs.~(U-B)$, $M_V~vs.~(B-V)$ , and $M_V~vs.~(V-I)$. Symbols have the same meaning as in Figs. \ref{figabsmagbc1} and \ref{figabsmagbc2}.}}
\end{figure*}

Figure \ref{figcolor} shows color-magnitude diagrams for the SN progenitor models discussed here. We can see that the progenitors from stars with higher initial mass ($\mini > 20 \msun$) are blue and have  negative $U-B$, $B-V$, and $V-I$ colors, since these are LBVs, WN8s, and WOs, all with relatively high values of $\teff$. They are well separated in color from the SN progenitors from stars with $\mini < 20 \msun$, which are either  RSGs and thus present red colors, or are YHGs, filling the gap between red and blue stars.

\begin{table*}
\begin{minipage}{\textwidth}
\scriptsize
\caption{Absolute magnitudes of the SN progenitor models in different filters: Johnson-Cousins $UBVRI$,  {\it HST}/WFPC2 F170, F336W, F450W, F555W, F606W, and F814W, and 2MASS $JHK_S$.}
\label{absmagbc1}
\centering
\vspace{0.1cm}
\begin{tabular}{l c c c c c c c c c c c  c c c c}
\hline\hline
M$_\mathrm{ini}$&  Sp. Type & $M_U$& $M_B$ & $M_V$ & $M_R$ & $M_I$ & $M_{170}$ & $M_{336}$  & $M_{450}$ & $M_{555}$ & $M_{606}$ & $M_{814}$ & $M_J$& $M_H$ & $M_K$ \\ 
(\msun) & & (mag) & (mag)& (mag)& (mag)& (mag)& (mag)& (mag)& (mag)& (mag)& (mag)& (mag) & (mag)& (mag)& (mag)
\\  \hline
 \hline
 \multicolumn{16}{c}{non-rotating models }\\
 \hline 
          9 &  $K-M~ I$ &  -0.67 &  -2.65 &  -4.21 &  -5.41 &  -6.98 &  -1.77 &  -0.43 &  -3.04 &  -4.19 &  -4.79 &  -6.92&  -8.51 &  -9.43 &  -9.71 \\
          12 & $K-M~I$ &  -0.95 &  -3.16 &  -4.81 &  -5.90 &  -7.25 &  -2.08 &  -0.73 &  -3.57 &  -4.81 &  -5.38 &  -7.19&  -8.66 &  -9.56 &  -9.82 \\
          15 & $K-M~I$ &  -1.57 &  -3.82 &  -5.49 &  -6.56 &  -7.86 &  -2.70 &  -1.36 &  -4.24 &  -5.49 &  -6.06 &  -7.81&  -9.26 & -10.14 & -10.40 \\
          20 & $K-M~I$ &  -2.54 &  -5.02 &  -6.76 &  -7.69 &  -8.67 &  -3.60 &  -2.27 &  -5.46 &  -6.77 &  -7.30 &  -8.63&  -9.92 & -10.73 & -10.97 \\
          23 & $K-M~I$ &  -4.25 &  -6.41 &  -7.90 &  -8.56 &  -9.14 &  -4.28 &  -3.47 &  -6.84 &  -7.90 &  -8.30 &  -9.13& -10.07 & -10.63 & -10.78 \\
          25 & WN11h/LBV &  -7.47 &  -6.38 &  -6.17 &  -6.14 &  -5.99 &  -8.76 &  -7.95 &  -6.38 &  -6.22 &  -6.20 &  -6.01&  -5.81 &  -5.75 &  -5.85 \\
          32 & WN7--8o &  -6.76 &  -5.61 &  -5.48 &  -5.45 &  -5.47 &  -8.67 &  -7.27 &  -5.62 &  -5.51 &  -5.50 &  -5.48&  -5.45 &  -5.55 &  -5.70 \\
          40 & WN7--8o &  -7.09 &  -6.11 &  -5.85 &  -5.91 &  -5.86 &  -8.98 &  -7.61 &  -6.12 &  -5.97 &  -5.93 &  -5.91&  -5.91 &  -5.98 &  -6.14 \\
          50 & WO1--3 &  -4.30 &  -3.15 &  -3.07 &  -3.09 &  -3.19 &  -6.17 &  -4.68 &  -3.20 &  -3.13 &  -3.10 &  -3.19&  -3.35 &  -3.58 &  -3.83 \\
          60 & WO1--3 &  -3.79 &  -2.69 &  -2.58 &  -2.61 &  -2.71 &  -5.84 &  -4.22 &  -2.74 &  -2.66 &  -2.62 &  -2.71&  -2.85 &  -3.07 &  -3.29 \\
          85 & WO1--3 &  -4.29 &  -3.18 &  -3.09 &  -3.12 &  -3.22 &  -6.31 &  -4.73 &  -3.23 &  -3.16 &  -3.13 &  -3.21&  -3.37 &  -3.59 &  -3.82 \\
         120 & WO1--3 &  -6.20 &  -4.62 &  -4.15 &  -4.14 &  -4.20 &  -7.65 &  -5.89 &  -4.70 &  -4.26 &  -4.18 &  -4.19&  -4.29 &  -4.38 &  -4.48 \\

  \hline
 \multicolumn{16}{c}{rotating models, initial $\vrot/\vcrit=0.4$ }\\
 \hline 
           9 & $K-M~I$ &  -0.83 &  -2.95 &  -4.57 &  -5.70 &  -7.13 &  -1.95 &  -0.60 &  -3.36 &  -4.56 &  -5.14 &  -7.07&  -8.59 &  -9.49 &  -9.76 \\
          12 &  $K-M~I$ &  -1.44 &  -3.62 &  -5.27 &  -6.36 &  -7.73 &  -2.56 &  -1.21 &  -4.03 &  -5.26 &  -5.84 &  -7.68&  -9.16 & -10.05 & -10.31 \\
          15 &  $K-M~I$   &  -2.11 &  -4.46 &  -6.17 &  -7.18 &  -8.36 &  -3.23 &  -1.89 &  -4.89 &  -6.18 &  -6.73 &  -8.32&  -9.70 & -10.56 & -10.81 \\
          16 &  $ K-M~I $ &  -2.19 &  -4.56 &  -6.28 &  -7.28 &  -8.42 &  -3.30 &  -1.96 &  -5.00 &  -6.29 &  -6.84 &  -8.37&  -9.74 & -10.59 & -10.84 \\
          18 & $G1 Ia^+$ &  -6.05 &  -7.11 &  -8.11 &  -8.56 &  -8.97 &  -4.34 &  -5.14 &  -7.41 &  -8.11 &  -8.39 &  -8.96&  -9.52 &  -9.83 &  -9.91 \\
          20 &  LBV &  -7.92 &  -6.90 &  -6.76 &  -6.79 &  -6.71 &  -8.82 &  -8.37 &  -6.92 &  -6.81 &  -6.83 &  -6.73&  -6.67 &  -6.72 &  -6.90 \\
          25 &  LBV  &  -8.30 &  -7.26 &  -7.15 &  -7.24 &  -7.13 &  -9.32 &  -8.74 &  -7.30 &  -7.21 &  -7.25 &  -7.17&  -7.24 &  -7.32 &  -7.64 \\
          28 &  WN10-11 &  -7.83 &  -6.72 &  -6.54 &  -6.56 &  -6.42 &  -8.92 &  -8.31 &  -6.73 &  -6.59 &  -6.58 &  -6.47&  -6.38 &  -6.38 &  -6.68 \\
          32 &  WO1 &  -4.49 &  -3.01 &  -2.57 &  -2.59 &  -2.68 &  -6.08 &  -4.27 &  -3.09 &  -2.72 &  -2.61 &  -2.67&  -2.78 &  -2.87 &  -2.95 \\
          40 &  WO1--2 &  -3.81 &  -2.69 &  -2.56 &  -2.59 &  -2.69 &  -5.86 &  -4.21 &  -2.74 &  -2.65 &  -2.60 &  -2.68&  -2.81 &  -3.02 &  -3.23 \\
          60 &  WO1--3 &  -4.09 &  -3.00 &  -2.92 &  -2.95 &  -3.06 &  -6.11 &  -4.55 &  -3.05 &  -2.99 &  -2.96 &  -3.05&  -3.22 &  -3.44 &  -3.68 \\
          85 &  WO1--3 &  -4.55 &  -3.47 &  -3.39 &  -3.43 &  -3.54 &  -6.53 &  -5.00 &  -3.51 &  -3.46 &  -3.44 &  -3.54&  -3.72 &  -3.95 &  -4.20 \\
         120 &  WO1 &  -4.09 &  -3.00 &  -2.92 &  -2.95 &  -3.06 &  -6.11 &  -4.55 &  -3.05 &  -2.99 &  -2.96 &  -3.05&  -3.22 &  -3.44 &  -3.68 \\

         \hline
\end{tabular}
\end{minipage}
\end{table*}

\begin{table*}
\begin{minipage}{\textwidth}
\scriptsize
\caption{Bolometric corrections of the SN progenitor models in different filters: Johnson-Cousins $UBVRI$,  {\it HST}/WFPC2 F170, F336W, F450W, F555W, F606W, and F814W, and 2MASS $JHK_S$.}
\label{absmagbc2}
\centering
\vspace{0.1cm}
\begin{tabular}{l c c c c c c c c c c c  c c c c}
\hline\hline
M$_\mathrm{ini}$&  Sp. Type & $BC_U$& $BC_B$ & $BC_V$ & $BC_R$ & $BC_I$ & $BC_{170}$ & $BC_{336}$  & $BC_{450}$ & $BC_{555}$ & $BC_{606}$ & $BC_{814}$ & $BC_J$& $BC_H$ & $BC_K$ \\ 
(\msun) & & (mag) & (mag)& (mag)& (mag)& (mag)& (mag)& (mag)& (mag)& (mag)& (mag)& (mag) & (mag)& (mag)& (mag)
\\  \hline
 \hline
 \multicolumn{16}{c}{non-rotating models }\\
 \hline 
           9 &  $K-M~ I$ &  -6.08 &  -4.10 &  -2.54 &  -1.34 &   0.23 &  -4.98 &  -6.32 &  -3.71 &  -2.56 &  -1.96 &   0.17&   1.76 &   2.68 &   2.96 \\
          12 & $K-M~I$ &  -5.97 &  -3.77 &  -2.11 &  -1.02 &   0.32 &  -4.84 &  -6.19 &  -3.35 &  -2.12 &  -1.54 &   0.27&   1.74 &   2.63 &   2.90 \\
          15 & $K-M~I$ &  -5.95 &  -3.70 &  -2.03 &  -0.96 &   0.34 &  -4.82 &  -6.16 &  -3.28 &  -2.03 &  -1.46 &   0.29&   1.74 &   2.62 &   2.88 \\
          20 & $K-M~I$ &  -5.68 &  -3.20 &  -1.46 &  -0.53 &   0.45 &  -4.61 &  -5.94 &  -2.75 &  -1.45 &  -0.92 &   0.42&   1.70 &   2.51 &   2.75 \\
          23 & $K-M~I$ &  -4.24 &  -2.08 &  -0.59 &   0.07 &   0.66 &  -4.20 &  -5.02 &  -1.64 &  -0.59 &  -0.19 &   0.64&   1.59 &   2.14 &   2.30 \\
          25 & WN11h/LBV &  -1.23 &  -2.33 &  -2.53 &  -2.57 &  -2.72 &   0.05 &  -0.75 &  -2.32 &  -2.48 &  -2.50 &  -2.70&  -2.90 &  -2.96 &  -2.85 \\
          32 & WN7--8o &  -2.42 &  -3.57 &  -3.70 &  -3.72 &  -3.70 &  -0.51 &  -1.90 &  -3.55 &  -3.66 &  -3.67 &  -3.70&  -3.72 &  -3.63 &  -3.48 \\
          40 & WN7--8o &  -2.38 &  -3.36 &  -3.62 &  -3.56 &  -3.61 &  -0.49 &  -1.86 &  -3.35 &  -3.50 &  -3.54 &  -3.56&  -3.56 &  -3.49 &  -3.33 \\
          50 & WO1--3 &  -5.64 &  -6.78 &  -6.87 &  -6.85 &  -6.75 &  -3.76 &  -5.25 &  -6.74 &  -6.81 &  -6.83 &  -6.75&  -6.59 &  -6.36 &  -6.11 \\
          60 & WO1--3 &  -5.68 &  -6.78 &  -6.89 &  -6.86 &  -6.76 &  -3.63 &  -5.25 &  -6.73 &  -6.81 &  -6.85 &  -6.76&  -6.62 &  -6.40 &  -6.18 \\
          85 & WO1--3 &  -5.85 &  -6.96 &  -7.06 &  -7.03 &  -6.93 &  -3.84 &  -5.42 &  -6.91 &  -6.99 &  -7.01 &  -6.93&  -6.78 &  -6.56 &  -6.32 \\
         120 & WO1--3 &  -4.69 &  -6.26 &  -6.74 &  -6.75 &  -6.69 &  -3.23 &  -5.00 &  -6.19 &  -6.63 &  -6.71 &  -6.70&  -6.60 &  -6.51 &  -6.41 \\
  \hline
 \multicolumn{16}{c}{rotating models, initial $\vrot/\vcrit=0.4$ }\\
 \hline 
           9 & $K-M~I$ &  -6.01 &  -3.89 &  -2.27 &  -1.14 &   0.29 &  -4.89 &  -6.24 &  -3.48 &  -2.28 &  -1.69 &   0.23&   1.75 &   2.65 &   2.92 \\
          12 &  $K-M~I$ &  -5.98 &  -3.79 &  -2.14 &  -1.05 &   0.32 &  -4.85 &  -6.20 &  -3.38 &  -2.15 &  -1.57 &   0.26&   1.74 &   2.64 &   2.90 \\
          15 &  $K-M~I$   &  -5.87 &  -3.52 &  -1.80 &  -0.79 &   0.39 &  -4.74 &  -6.08 &  -3.08 &  -1.80 &  -1.24 &   0.34&   1.72 &   2.59 &   2.84 \\
          16 &  $ K-M~I $ &  -5.83 &  -3.46 &  -1.73 &  -0.74 &   0.40 &  -4.72 &  -6.06 &  -3.02 &  -1.73 &  -1.18 &   0.36&   1.72 &   2.58 &   2.82 \\
          18 & $G1 Ia^+$ &  -2.17 &  -1.11 &  -0.11 &   0.34 &   0.74 &  -3.89 &  -3.08 &  -0.81 &  -0.11 &   0.16 &   0.74&   1.30 &   1.60 &   1.68 \\
          20 &  LBV &  -0.55 &  -1.57 &  -1.70 &  -1.67 &  -1.75 &   0.35 &  -0.10 &  -1.54 &  -1.65 &  -1.63 &  -1.73&  -1.79 &  -1.74 &  -1.56 \\
          25 &  LBV  &  -0.72 &  -1.75 &  -1.87 &  -1.77 &  -1.88 &   0.30 &  -0.27 &  -1.72 &  -1.81 &  -1.77 &  -1.84&  -1.78 &  -1.70 &  -1.38 \\
          28 &  WN10-11 &  -1.28 &  -2.40 &  -2.58 &  -2.56 &  -2.70 &  -0.20 &  -0.81 &  -2.39 &  -2.53 &  -2.54 &  -2.65&  -2.74 &  -2.74 &  -2.44 \\
          32 &  WO1 &  -4.59 &  -6.07 &  -6.51 &  -6.49 &  -6.40 &  -3.00 &  -4.81 &  -5.99 &  -6.36 &  -6.47 &  -6.41&  -6.30 &  -6.21 &  -6.13 \\
          40 &  WO1--2 &  -5.55 &  -6.68 &  -6.81 &  -6.78 &  -6.68 &  -3.51 &  -5.15 &  -6.62 &  -6.72 &  -6.76 &  -6.68&  -6.56 &  -6.35 &  -6.14 \\
          60 &  WO1--3 &  -5.90 &  -6.99 &  -7.08 &  -7.04 &  -6.94 &  -3.88 &  -5.45 &  -6.95 &  -7.01 &  -7.03 &  -6.94&  -6.78 &  -6.56 &  -6.31 \\
          85 &  WO1--3 &  -6.01 &  -7.09 &  -7.16 &  -7.12 &  -7.01 &  -4.02 &  -5.55 &  -7.04 &  -7.10 &  -7.11 &  -7.01&  -6.84 &  -6.61 &  -6.36 \\
         120 &  WO1 &  -6.00 &  -7.09 &  -7.17 &  -7.14 &  -7.03 &  -3.98 &  -5.54 &  -7.04 &  -7.10 &  -7.13 &  -7.04&  -6.87 &  -6.65 &  -6.41 \\
         \hline
\end{tabular}
\end{minipage}
\end{table*}

\section{\label{detect} Detectability of core-collapse SN progenitors}

\subsection{Progenitors of SN IIP}
Volume-limited SN surveys show that SN IIP should be the most common type of core-collapse SN, corresponding to about 48--55\% of the total of core-collapse SNe \citep{smith11sn,eldridge13}. Our results show that the progenitors of SN IIP (triangles in Fig. \ref{figabsmagbc1}) are the faintest progenitors in the $F170W$ and $U$ filters, becoming brighter in the red filters, and are the brightest ones in the $IJHK_S$ filters.  This is because they are RSGs, which have the peak of their flux emitted in the $IJHK_S$ filters (Fig. \ref{snprogsed}). Starting from the $R$-band, even the faintest SN IIP progenitors have $\mbol < -5$~mag, and thus should be the easiest to be detected in $RIJHK_S$.

\begin{figure*}
\center
\resizebox{0.75\hsize}{!}{\includegraphics{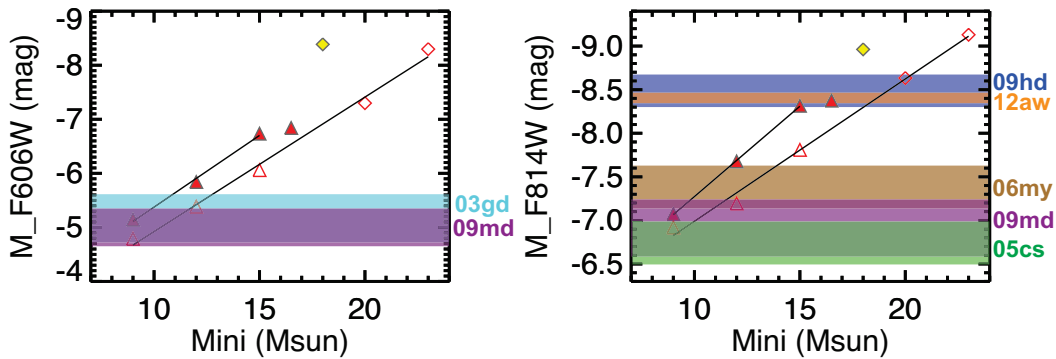}}
\caption{\label{magmassrsg} {Absolute magnitudes in the {\it HST}/WFPC F606W and F814W filters as a function of initial mass for RSGs at the pre-SN stage. Symbols are as in Fig. \ref{figabsmagbc1}. Least-squares linear fits for both models (see text) are also shown. The colored bands represent the absolute magnitude $\pm 1$ sigma of observed RSGs that underwent core-collapse SN.}}
\end{figure*}

As found in previous studies \citep[e.g.,][]{humphreys84,levesque05,davies13}, the absolute magnitudes of RSGs in a given filter significantly change as a function of temperature and luminosity. Since these are well correlated with the initial mass, one can use the absolute magnitudes of SN progenitors that were RSGs before the explosion to estimate their initial masses. Here we provide relationships between \mini\ and absolute magnitudes that should be applied to de-reddened photometry, keeping in mind that the reddening may be uncertain towards RSGs, which could affect the determination of the de-reddened photometry from the observations \citep{kochanek12,walmswell12,davies13}. Figure~\ref{magmassrsg} shows the variation of the absolute magnitude in different filters of the {\it HST}/WFPC2 system, which is commonly employed in studies of SN progenitors, as a function of initial mass for the models that are RSGs at their end stage. In most filters, one can clearly devise a linear variation of the absolute magnitude as a function of initial mass, both for rotating and non-rotating models. We notice, however, that the absolute magnitude of rotating models seem to present a plateau and level off above 15~\msun. Therefore, we computed the following least-square linear fits for rotating models, valid in the range $9\msun < \mini < 15\msun$:
\begin{equation}
\mini (\msun)= -3.977 -3.903 M_\mathrm{F450W} 
\end{equation}
\vspace{-0.5cm}
\begin{equation}
 \mini (\msun)= -10.165 -3.753 M_\mathrm{F606W}
\end{equation}
\vspace{-0.5cm}
\begin{equation}
 \mini (\msun)= -25.108 -4.827 M_\mathrm{F814W}
\end{equation}
 while for non-rotating models we found (valid in the range $9\msun < \mini (\msun) < 23\msun$):
\begin{equation}
 \mini (\msun)= -4.277 -4.480M_\mathrm{F450W}
\end{equation}
\vspace{-0.5cm}
\begin{equation}
 \mini (\msun)= -11.566 -4.346 M_\mathrm{F606W}
\end{equation}
\vspace{-0.5cm}
\begin{equation}
 \mini (\msun)= -32.822 -6.129 M_\mathrm{F814W}
\end{equation}

\begin{sidewaystable}
\begin{minipage}{\textwidth}
\scriptsize
\caption{Fundamental data of observed SN II progenitors with metallicity around solar ($\log \mathrm{(O/H)}+12\simeq8.4-8.7$).  All progenitors are from SN IIP, with the exception of the progenitor of SN 2009hd, which was a SN IIL. We quote the distance, reddening, absolute magnitudes in R, I, F555W, F606W, F814W, J, and K$_S$ filters, and initial progenitor mass found in the literature, together with  our determination of the progenitor's initial mass based on our rotating and non-rotating models.}
\label{snmass}
\centering
\vspace{0.1cm}
\begin{tabular}{l c c c c c c c  c  c c c c c c  }
\hline\hline
SN  & d &  $A_V$  & \magr & \magi &$M_\mathit{555}$ &$M_\mathit{606}$ & $M_\mathit{814}$ &$M_\mathit{J}$ &$M_\mathit{K_s}$ & M$_\mathrm{ini}$ lit. &  Refs. & M$_\mathrm{ini}$ non-rot. & M$_\mathrm{ini}$ rot.  \\  
       &     (Mpc)                    & (mag)   & (mag)   & (mag)    & (mag)                              & (mag)                                         & (mag)                   & (mag)              & (mag)                              &   (\msun)              &      &         (\msun) 				 &   (\msun) \\
\hline
1999br    &  14.1$\pm$ 2.6  & 0.06 $\pm$ 0.06  &$-$&$-$&$-$&    $>-5.88$ &$-$&$-$&$-$		&    $<15$    & MS05, S09b		 & $<14 $& $ <12$      \\
1999em  &  11.7 $\pm$ 1.0 & 0.31 $\pm$ 0.16  &$-$ &    $>-7.49$&$-$&$-$&$-$&$-$&$-$ 		&    $<15$	   & S09b 		                    & $<13 $& $ <11$     	 \\
1999gi   &   10.0 $\pm$ 0.8     &0.65 $\pm$ 0.16&$-$&$-$&$-$&    $>-5.67$ &$-$&$-$&$-$ 	          &    $<14$   & S01,S09b	     		 & $<13 $& $ <11$  	\\ 
2001du   & 18.3 $\pm$ 1.2 & 0.53 $\pm$ 0.28    &$-$&$-$&$-$&$-$ &$>-7.36$&$-$&$-$                  	&    $<15$	    & vD03, S03, S09b        	 & $<12 $& $ <10$ 	\\
2002hh   & $5.9 \pm 0.4$    & 5.2 $\pm$  0.2      &$-$ &    $>-8.15$&$-$&$-$&$-$&$-$&$-$ 	         &    $<18$   & S09b 		 		 & $<17 $& $ <14$ 		 \\ 
2003gd   & 9.3 $\pm$ 1.8    & 0.43 $\pm$ 0.19   &$-$ &    $-6.92 \pm 0.47$&$-$&   $-5.16 \pm 0.44$&$-$&$-$&$-$ & $8.4\pm2.0$ &S09b, M13   & $10.8\pm2.1$ & $8.5\pm1.8  $ 	 \\
2003ie   & 15.5 $\pm$ 1.2   &0.04 	    &    $>-8.33$ &$-$&$-$&$-$&$-$&$-$&$-$ 			&      $<25$    	     & S09b                     	 & $<23 $& $ <20$ 		      \\
2004dg    & $20.0 \pm 2.6$  & 0.74 $\pm$ 0.09     &$-$&$-$&$-$&$-$ &$>-6.93$&$-$&$-$            &    $<12$    	     & S09b                    	 & $<10 $& $ <8$ 		  \\
2005cs    & $8.4 \pm 1.0$     & 0.43 $\pm$ 0.06 &$-$&$-$&$-$&$-$& $-6.75 \pm 0.26$&$-$&$-$   &    $9.5^{+3.4}_{-2.2}$ &   M05,L06,E07,S09b     & $8.6\pm5.4$& $ 7.5\pm1.5$ \\
2006bc   & $14.7 \pm 2.6$    & 0.64	          &$-$&$-$&$-$&$-$ &$>-6.75$&$-$&$-$                              &    $<12$	    	& S09b	                    & $<8.5 $& $ <7.5$ 	   \\
2006my   & $22.3 \pm 2.6$    &0.08 	          &$-$&$-$&$-$&$-$& $-7.37 \pm 0.25$&$-$&$-$		&      $9.8\pm1.7$   & M13 		  & $12.4\pm3.5 $& $ 10.5\pm1.5$ 		    \\
2007aa   & $20.5 \pm 2.6$    &  0.09	    &$-$&$-$&$-$&$-$ &>-7.16&$-$&$-$           	&  $<12$    				 &S09b             	 & $<11 $& $ <9.5$ 		\\
2009md  & $21.3 \pm 2.1$    & 0.31   &$-$&$-$&$-$&$-5.00 \pm 0.35$& $-6.84\pm0.33$&$-$&$-$	& $8.5^{+6.5}_{-1.5}$ & F11 	& $10.0 \pm 1.8$ & $8.1\pm1.4$  \\
2009hd   & $9.4 \pm 0.4$       & $3.8 \pm 0.14 $ &$-$&$-$&$-$&$-$ &$>8.54$ &$-$&$-$   & $<20$	       & E11    & $<20$ & $<17$ \\
2012aw  & $10.0 \pm 0.4$   & 3.1 		&$-$&$-$&$-6.69 \pm 0.15$&$-$& $-8.39\pm0.11$&$-9.85 \pm 0.10$&$-10.88 \pm 0.21$  &$17-18$& vD12 & $18.9 \pm 1.8$ & $15.6\pm0.9$  \\
...  & ... & ...		&$-$&$-$&$-6.48 \pm 0.15$&$-$& $-8.39\pm0.11$&$-9.77 \pm 0.22$&$-11.2 \pm 0.4$  &$14-26$& F12 & $18.3 \pm 1.8$ & $15.5\pm0.9$  \\

\hline
\end{tabular}
\tablefoot{References: S01=\citet{smartt01}; S03=\citet{smartt03}; vD03=\citet{vandyk03c}; M05=\citet{maund05}; MS05=\citet{ms05}; L06=\citet{li06}; L07=\citet{li07}; E07=\citet{eldridge07}; S09b=\citet{smartt09b}; E11=\citet{eliasrosa11}; F11=\citet{fraser11}; F12=\citet{fraser12}; vD12=\citet{vandyk12_12aw}; M13=\citet{maund13a}. }
\end{minipage}
\end{sidewaystable}

It is readily apparent that rotating and non-rotating models follow different relationships. For the same mass, we find that rotating models are systematically brighter than non-rotating models, in all filters. This is due to the rotating models having higher $\teff$ and $\lstar$ than non-rotating models at the same mass. The higher $\lstar$ in rotating models is caused by the larger He cores in these models compared to non-rotating models.

Since it is impossible to distinguish, based on photometry only, whether a SN progenitor come from an initially rotating or non (slowly) rotating star, this implies a degeneracy between the absolute magnitude at the pre-SN stage and the initial rotation and initial mass of SN progenitors that were RSGs. Because of this degeneracy, there is an uncertainty in the predicted initial mass of the SN progenitor. When analyzing the progenitor of SN2012aw, \citet{vandyk12_12aw} reached a similar conclusion when comparing their results based on rotating models from \citet{ekstrom12} with those from \citet{fraser12}, which in turn had employed non-rotating STARS models \citep{eldridge08}.

Once the absolute magnitude in a given filter of a SN progenitor is known, its mass can be estimated by employing Equations 4--9, assuming it is a RSG. These relationships assume that stars have their terminus in the HR diagram with $\teff$ and $\lstar$ as predicted by our evolutionary models. Under this assumption, we are thus able to constrain the mass of observed SN progenitors in a homogeneous way, as shown in Fig. \ref{magmassrsg}. Table \ref{snmass} presents the assumed distance, reddening,  absolute magnitudes, and mass estimate of SN progenitors collected from the literature, together with our estimates. We consider here only progenitors that are in galaxies consistent with solar metallicity, with Oxygen abundance (by number) of $\log \mathrm{(O/H)}+12\simeq8.4-8.7$.

In the cases where a progenitor has been found, such as SN 2003gd \citep{smartt09b,maund13a}, SN 2005cs \citep{maund05,li06,eldridge07,smartt09b}, SN 2006my \citep{maund13a}, SN 2009hd \citep{eliasrosa11}, SN 2009md \citep{fraser11}, and 2012aw \citep{fraser12,vandyk12_12aw}, 
our estimates for the non-rotating and rotating models brackett well
the values derived by the Smartt group, which employ non-rotating STARS models.
In the cases where only an upper limit to the absolute magnitude of the SN progenitor is available, we derive in general lower values for the upper initial mass limit, both for non-rotating and rotating models.
This results from the fact that we assume that RSGs at their end phase have $\teff$ and $\lstar$ as predicted by our evolutionary models at the end of core-C burning (Table \ref{model log}). Conversely, while the Smartt group has also assumed that RSGs most likely explode at the end of their evolution, they adopted a conservative approach to compute the progenitor upper mass limit, taking into account that the star could actually have exploded at any point after the end of core-He burning. Also, because rotating models are brighter than non-rotating models of the same mass, our upper mass limit is lower for the rotating models.

\subsection{Progenitors of SN Ibc}

With our database of SN Ibc progenitor spectra and absolute magnitudes, we are able to assess their detectability with much more realism than before. Previous works employed empirical calibrations of absolute magnitudes and bolometric corrections \citep{yoon12,ms05,maund05b,crockett07}. More recently,  \citet{eldridge13} employed atmospheric models of WR stars from the Potsdam group \citep{hamann06,sander12} that were computed for a range of physical parameters, but not self-consistently with the evolutionary calculations. Here, our quantities are calculated self-consistently for each progenitor based on our combined stellar evolution and atmospheric modeling. However, we recall here the caveats associated with classifying the SN type based on the chemical composition of the ejecta, and the challenges of discriminating between SN Ib and Ic based on the He abundance \citep[e.g.][]{dessart11b}. The following discussion assumes that SN Ib have more than $0.6~\msun$ of He in their ejecta, while quantities smaller than this would lead to SN Ic.

So far, no WR star has been directly detected as a SN progenitor \citep{smartt09a,eldridge13}. Since single-star massive evolution models predict WRs as SN progenitors, the yet non-detection may sound surprising. It has been proposed that binary evolution could be absolutely needed to explain the non-detections \citep{smartt09a,eldridge13}. However, using empirical bolometric corrections based on WR models from the Potsdam group, \citet{yoon12} suggested that at their end stage, single WR stars have faint absolute magnitudes in the optical bands, since they are extremely hot ($\log (\teff / \mathrm{K}) > 5.0$). This would make them undetectable in the optical pre-explosion images of SN Ibc progenitors.

Here, we showed more specifically that the progenitors of SN Ibc are WO 1--3, WN7--8, WN10--11 stars, and computed their absolute magnitudes and bolometric corrections. We find that, contrary to RSGs, the absolute magnitudes of the WOs that are SN progenitors do not show a clear scaling with initial mass. This happens because the models with higher \mini\ finish at higher luminosities but also at  hotter temperatures. This increase in $\teff$ causes more flux to be emitted outside the optical and near-infrared bands, which roughly compensates the increase in flux caused by the higher luminosity. Therefore, when photometric observations of SN Ibc detect their progenitors, our models predict that it will be challenging to assign a stellar mass with similar precision as has been done for RSGs.

\begin{figure}
\resizebox{0.99\hsize}{!}{\includegraphics{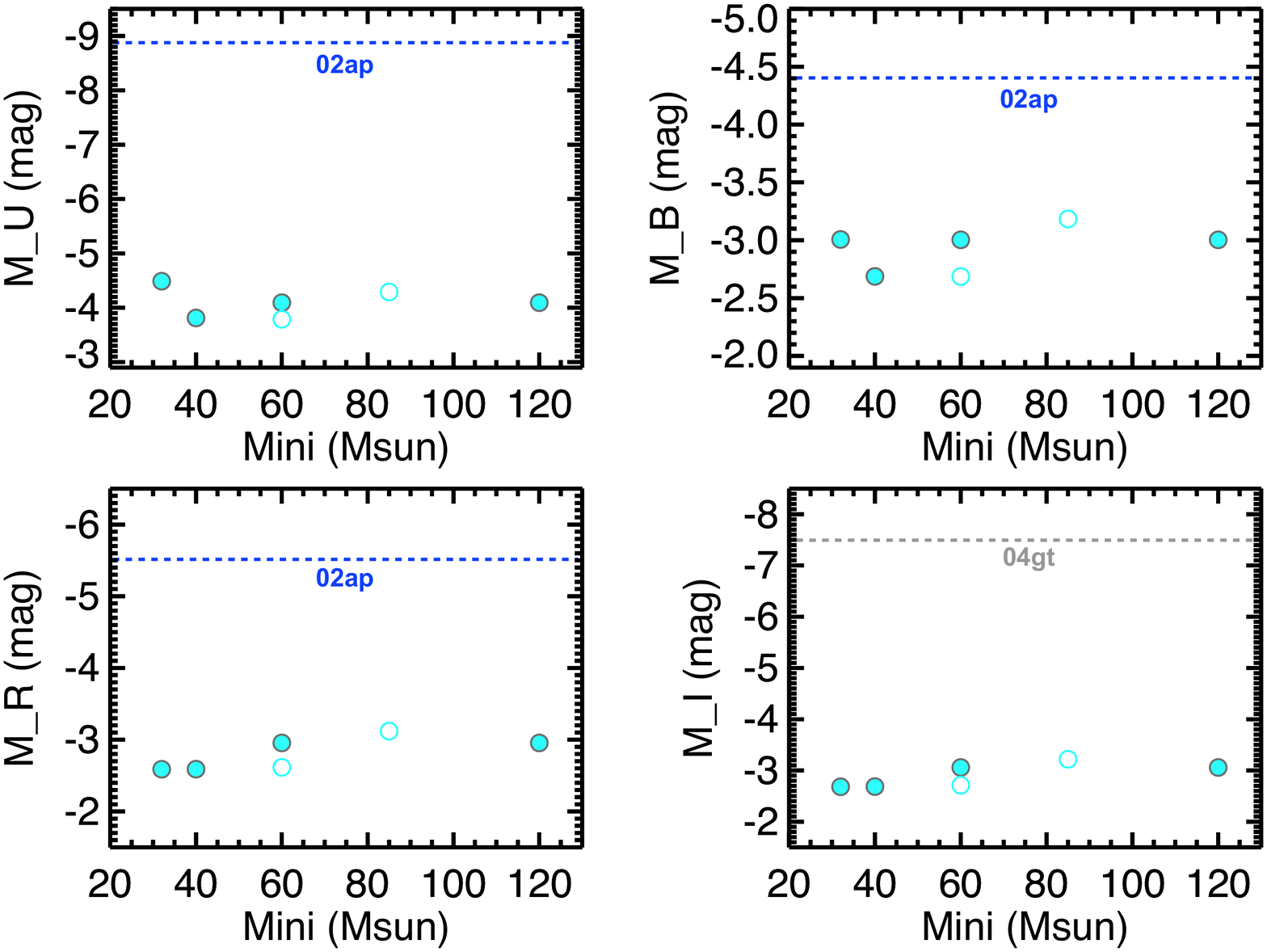}}\\
\resizebox{0.99\hsize}{!}{\includegraphics{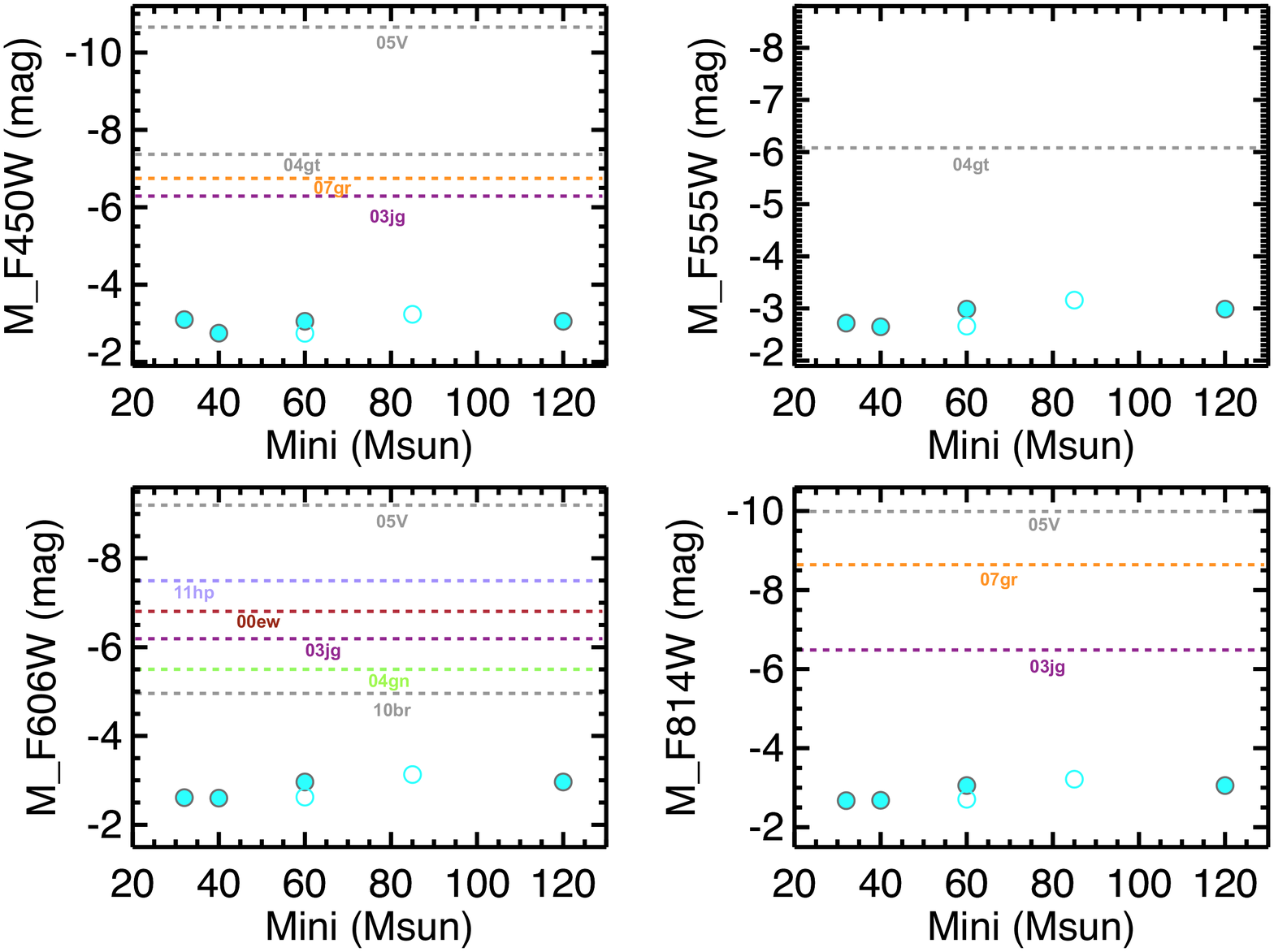}}
\caption{\label{figsnicdetect} {Absolute magnitude of our models that are SN Ic progenitors (open and filled cyan circles) compared to observed upper magnitude limit for different SNe Ic (colored horizontal dashed lines followed by shortened SN label) and SN Ibc (gray horizontal dashed lines). Open (filled) symbols correspond to non-rotating (rotating) models. From top to bottom, we show the comparison between model and observations in the $UBRI$ and {\it HST}/WFPC2 F450W, F555W, F606W, and F814W filters. Note that all models lie below the detectability limit of all SN Ic and Ibc. }}
\end{figure}

Let us investigate whether our SN Ib and SN Ic model progenitors would be detectable in the available observational data. We adopt the observed magnitude limits compiled by \citet{eldridge13} of progenitors of 6 SN Ic, 3 SN Ib, and 3 SN Ibc. In this context, a SN Ibc classification  means that the distinction between a Ib or Ic classification is unclear. Figure \ref{figsnicdetect} presents the observed detection limits in the absolute magnitude of SN Ic progenitors compared to the predictions of our models in different filters. We find that all SN Ic model progenitors would be undetectable in the available pre-explosion images. Included in Fig. \ref{figsnicdetect} are also SN Ibc progenitors, showing that if these were SN Ic, their progenitors would be undetectable too. In the case of the SN Ic progenitor with the deepest magnitude limit available ($M_B\simeq-4.4$mag, SN 2002ap), our models predict that even the brightest SN Ic progenitor would still be 1.5 mag fainter than the detection limit.  Our conclusion is that SN Ic progenitors from single stars can easily evade detection in pre-explosion images, since they have spectral type WO and are faint in the optical/NIR bands. This is because they are extremely hot at their endpoints, in agreement with the suggestion from \citet{yoon12}. {\it Thus, our models of single stars are in agreement with the lack of detection of SN Ib and Ic progenitors at the current magnitude limit of the observations.}

Our results are in contrast with the findings of \citet{eldridge13}, who find a 13\% of probability of non-detection of the progenitors of SN Ib and Ic based on single star evolution models, and 12\% non-detection probability based on binary evolution models. The reason for this discrepancy is that we assume, as in \citet{yoon12}, that stars will explode only at the end point of the evolution predicted by stellar evolution models. \citet{eldridge13}, on the contrary, assumes that the stars can explode anytime after the end of core-helium burning. Our models predict that there is significant increase in the effective temperature of the star after the end of core-helium burning, so assuming an explosion before the end point of the evolution would underestimate the temperature of the progenitor, and overestimate its absolute magnitude. Therefore, our models predict much fainter progenitors than those from \citet{eldridge13}.

One may wonder what would be the maximum distance that a SN Ic progenitor should be to be still detectable according to our predicted absolute magnitudes. This obviously depends on the magnitude limits and the amount of reddening towards the progenitor. For typical detection limits ($m=24.5$) and low amounts of extinction (0.3 mag in a given band), this implies a maximum distance to SN Ic progenitors of 2.7 Mpc for detection. For a SN 2002ap--like detection limit ($m_B=26.0$~mag, \citealt{crockett07}), the maximum distance would be 5.5~Mpc. Obviously, these are upper limits that would decrease if the progenitor is seen under large extinction.  We predict that the limiting magnitudes in the observations should be at least 2 mag fainter than the current best limit in the $B$ band to detect single star progenitors of SN Ic at distances of 10--20~Mpc.

\begin{figure}
\resizebox{0.99\hsize}{!}{\includegraphics{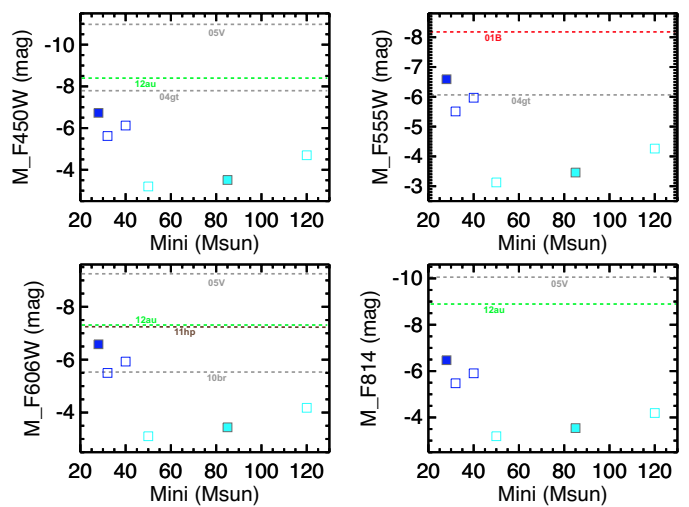}}
\caption{\label{figsnibdetect} {Similar to Fig. \ref{figsnicdetect}, but for SN Ib progenitors. }}
\end{figure}

The absolute magnitude of SN Ib progenitors from our models and the detection limits derived from observations are shown in Fig. \ref{figsnibdetect}. In all filters, we see that the three observed magnitude limits of SN Ib progenitors are brighter than the absolute magnitude predicted by our models of SN Ib progenitors. Therefore, according to our models, the progenitors of SN 2001B, SN 2011am and SN 2012au should not have been detected, in agreement with the observations. 

The magnitude limits of the progenitors of the SN Ibc SN 2004gt and SN 2010br are also show in Fig. \ref{figsnibdetect}. As mentioned above, these are SNe where the classification is uncertain between Ib and Ic. Fig. \ref{figsnibdetect} reveals that the progenitors of SN 2004gt and SN 2010br could have been detected in pre-explosion images if they were progenitors of SN Ib. Since our models predict some of the SN Ib progenitors to be much brighter than the SN Ic progenitors, to determine the detectability of the progenitor it is thus of crucial importance to constrain the SN type between Ib and Ic. To asses the detectablility of the progenitors of SN 2004gt and SN 2010br, we compute the probability that a SN progenitor would {\it not} be detected in available pre-explosion images. For this purpose, we assume that the progenitor is randomly drawn from the mass range that produces a SN Ib or Ic, a Salpeter IMF, the initial mass ranges that generates SN Ibc from \citet{georgy12a}, and that half of the stars evolve from rotating stars, while the other half evolve from slow-rotating stars (thus described by our non-rotating models). We also linearly interpolate the synthetic model absolute magnitudes as a function of initial mass. We obtain that the progenitors of SN2004gt  and SN 2010br have a probability of non-detection in the pre-explosion images of 87\% and 62\%, respectively. As in the case of SN 2002ap, our non-detection probabilities are higher than those computed by \citet{eldridge13}, and the reasons are the same as for SN 2002ap.

All in all, we see that the current non-detection of progenitors of SN Ib and SN Ic in pre-explosion images cannot be used as an argument to discard single stars as the progenitors of these types of SNe. The present discussion also shows that
to really discard the single star scenarios for a given progenitor, no detection should be made even when the limiting
magnitudes in the observations would be at least 2 mag fainter than the current best limit in the B  band. Therefore, if our models are correct, binarity is not necessarily needed to explain the non-detections.

\section{WO stars as progenitors of SN Ibc}
\label{wosn}
Our results present a paradigm shift in the sense that for the most massive models, we find a WO spectral type before core collapse, and  {\it not} a WC spectral type as has been widely assumed based on the chemical composition of the progenitor \citep{georgy09,georgy12a,eldridge08}.  But is this shift in paradigm supported by observations? Based on the spectrum, our models indicate that the WO phase is a short evolutionary stage at the end, with duration of a few $\sim10000$ years (Groh et al. 2013, in preparation). The WC phase, on the other hand, corresponds to where the star spends most of its He burning phase, and thus has a relatively long duration. Therefore, one would expect that WC stars are a large fraction of the total number of WR stars, while WO stars should be extremely rare. Indeed, out of a population of 576 Galactic WR stars\footnote{According to the latest compilation of WR stars by P. Crowther, v1.5, Feb 2013; \url{http://pacrowther.staff.shef.ac.uk/WRcat}}, only four WO stars are known in the Galaxy (one of which is in a binary system). From a zeroth-order calculation of the expected ratio WO/WR based on our theoretical models and the predicted timescales (Groh et al. 2013, in preparation), one would expect 0.5 to 2\% of the total WRs to be WOs. This is in line with the observed rarity of WO stars.

\begin{figure}
\center
\resizebox{0.99\hsize}{!}{\includegraphics{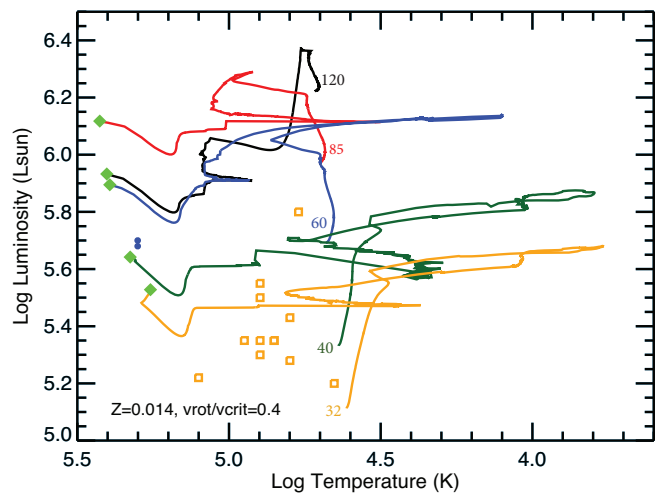}}
\caption{\label{wohrd} Evolutionary tracks of rotating models from  \cite{ekstrom12}, as those in Fig. \ref{hrd1}, with the endpoints indicated by green diamonds. The position of Galactic WO stars (filled blue circles) and WC stars with known distances (open orange squares) are overplotted for comparison. }
\end{figure}

We plot the location of Galactic WO and WC stars in the HR diagram in Fig. \ref{wohrd}, assuming the parameters determined by \citet{sander12} and restraining the analysis to the objects with known distances. For simplicity, we plot their \tstar\ (computed at a Rosseland optical depth of 20) as corresponding to the values of \tstar from our stellar evolution models. As discussed by  \citet{sander12}, this assumption could lead to uncertainties of 0.1 -- 0.2 dex in \tstar, which would not affect our conclusions below. We can immediately notice in Fig. \ref{wohrd} that the two observed Galactic WO stars, WR 102 and 142 (filled blue circles), lie extremely close to the endpoints of the rotating 40~\msun\ track (and to the non-rotating $60~\msun$ track, not shown in Fig.~\ref{wohrd}; see Fig. \ref{hrd1}). This would mean that their initial masses were around $40-60~\msun$, while errors in the \lstar\ and/or \tstar\ determination would shift the initial mass between 32-120~\msun, but still putting them close to the endpoint of the evolution. Therefore, the position of observed Galactic WO stars is consistent with our model results, which show that WO stars are the immediate stage before core collapse. In addition, the chemical surface abundances predicted by our models are similar to those derived for WOs by \citet{sander12}, with a high C and O content. We can say that our rotating models are corroborated by observations in the range $32-60~\msun$. \citet{sander12} reached a similar conclusion by comparing their observations with the \citet{meynet03} non-rotating evolutionary tracks.

No WO star with $\log \lstar/\lsun > 5.7$ has been detected in the Galaxy. In the HR diagram, these more luminous WOs would be in a place close to the endstage of the tracks with $\mini > 60~\msun$. One may then rightfully wonder whether stars born more massive than $60~\msun$ indeed end their lives as WO stars. We point out that the non-detection of WOs could be due to the short timescale related to this phase (5000--10000~years), which would make the detection quite unlikely. This would be combined with the scarcity of stars above $\mini > 60~\msun$, as for a Salpeter initial mass function, one would expect approximately twice as few stars between $60-120~\msun$ than in the $32-60~\msun$ range. Finally, as shown in Figs. \ref{figspec1} and \ref{figspec2}, the detection of WO stars may be hampered by the possibly weak lined-spectrum that could be characteristic of these stars, depending on their $\mdot$. Since most WR stars are identified based on narrow-band surveys, one would find challenging to detect weak-lined WO stars in narrow-band surveys using the \ion{C}{iv} $\lambda5808$ line, for instance. Our conclusion is that the current non-detection of WO stars with $\log \lstar/\lsun > 5.7$ is expected given their rarity.

An additional support for WO stars being the endstage of single massive stars in the range $32-120~\msun$ comes from the fact that the observed WC stars with known distances (orange squares in Fig. \ref{wohrd}) are located significantly far from the endpoint of the stellar evolutionary tracks. However, due to the fact that WC stars may have inflated radii  \citep{grafener12a}, it might be that, once corrected for this inflation effect, the stars would actually  populate the region near the end points of the evolutionary tracks (for recent discussions see \citealt{sander12} and \citet{georgy12a}). We do not think that this effect could shift sufficiently to the blue the observed positions of the WC stars. This is because at the endmost stages, $\tstar>150000$, which is the characteristic temperature of the opacity peak due to Fe. As shown by \citet{grafener12a}, one needs $\tstar<150000$ in order for inflation to occur, as the Fe opacity bump region needs to be within the envelope of the star, and not in the stellar wind as it occurs when $\tstar>150000$.

Our models also show that WO stars arise naturally at solar metallicity at the end stage of the evolution. Therefore, lower metallicities are not necessary to produce WO stars as SN Ibc progenitors. Also, WO stars occur both in rotating and non-rotating models, with the difference being that rotation diminishes the minimum initial mass to produce WO stars from $60~\msun$ to $32~\msun$. As such, WO stars do not come necessarily only from rapid rotators.

As for GRB progenitors, we note that GRBs seem to occur, albeit at lower frequency, even at metallicities around the solar values \citep{levesque10}. \citet{georgy12a} showed that 40 and 60~\msun\ rotating models discussed here have a large amount of core angular momentum, which seems to make them favorable for GRB production. If this turns out to indeed occur, our results here indicate that the GRB progenitors at solar metallicity have a WO spectral type.

\section{When LBVs are SN progenitors and their SN types}
\label{lbvssn}
From a theoretical perspective, LBVs have been only recently linked to the pre-SN stage of massive stars \citep{gme13}. In that paper, we showed that stars with $\mini=20-25~\msun$ end their lives with properties similar to LBVs, such as the spectral morphology, the proximity to the Eddington limit, and the chemical abundances. 

In addition, as can be inferred from their values of  \teff\ ($\sim20000~\K$), we note here that our $\mini=20-25~\msun$ rotating models flirt with the so-called bistability limit of line-driven winds. We refer the reader to \citet{pauldrach90}, \citet{lamers95}, and \citet{vink99} for detailed discussions. In short, the bistability is related to a change in the optical depth of the Lyman continuum, which in turn changes the ionization structure of metals in the wind, in particular Fe. This occurs at $\teff \sim21000~\K$. As a consequence, the amount of line driving is affected, ultimately changing $\mdot$ and \vinf.  Classical LBVs such as AG Car \citep{vink02,ghd09,ghd11} and P Cygni \citep{pauldrach90,najarro97,najarro01} are well known for being close to the bistability limit.  In the context of our models, being close to the bistability limit may produce an inhomogeneous circumstellar medium (CSM) in the vicinity of the progenitor (Moryia et al., in prep.). It is interesting to note that the inference of wind variability from the radio lightcurve was the first way to link LBVs as SN progenitors \citep{kv06}, and theoretical work has shown that LBVs produce inhomogeneous CSM when they cross the bistability \citep{gv11}.

\begin{figure}
\center
\resizebox{0.99\hsize}{!}{\includegraphics{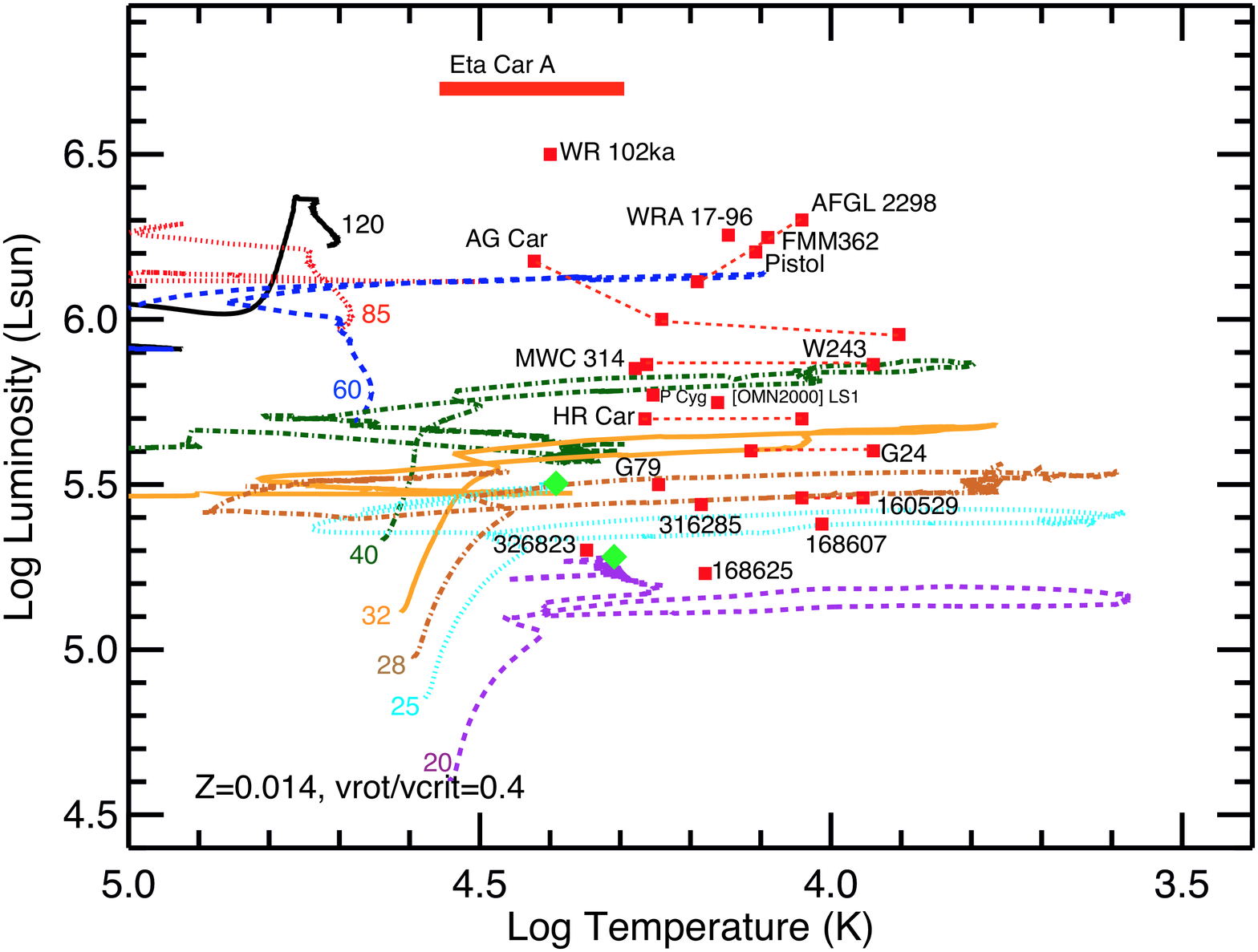}}
\caption{\label{lbvhrd} Position of Galactic LBV stars (red squares) in the HR diagram. The sources of the stellar parameters of the LBVs are as follows: Eta Car A \citep{ghm12,hillier01}, WR 102ka \citep{barniske08},  AG Car \citep[][Groh et al., in prep.]{ghd09,ghd11}, AFGL 2298 \citep{clark09}, FMM 362 and Pistol star \citep{najarro09}, Wray 17-96 \citep{egan02}, G24.73+0.69 \citep{clark09}, [OMN2000] LS1 \citep{clark09b}, HD 316285 \citep{hillier98}, HD 160529, HD 168625 and HD 168607 \citep{vg01}, P Cygni \citep[][Najarro 2011, priv. comm.]{najarro01}, HD 326823 \citep{marcolino07}, G79.29+0.46 \citep{trams98}, and W243 \citep{ritchie09}. For comparison, we show evolutionary tracks of rotating models from  \cite{ekstrom12}, as those in Fig. \ref{hrd1}, with the endpoints indicated by green diamonds. }
\end{figure}

Are all LBVs at the end stage of massive star evolution, or are some LBVs in a transitional stage towards the WR phase? This is a fair question that should be posed, in particular if one's goal is to obtain the evolutionary status of observed LBVs. Here we tackle this question from a theoretical perspective. For that purpose, we plot in Fig. \ref{lbvhrd} the position of the Galactic LBVs and LBV candidates in the HR diagram, together with our evolutionary tracks for rotating stars.  We immediately note in Fig. \ref{lbvhrd} that LBVs are distributed over a range of luminosities (see also \citealt{vg01,smithvink04,clark05,clark09,clark12b,vink12lbvs}). If LBVs arise from the evolution of single stars, the spread in luminosity means that they come from a range of initial masses. According to our models, LBVs would arise from stars with $\mini \geq 20~\msun$. We predict that {\it only} the 20-25~\msun\ tracks have their pre-SN stage as LBVs, and all LBVs that arise from stars with $\mini > 25~\msun$ will further evolve to the WR stage.

It is important to point out that, contrary to previous studies \citep{smithvink04}, there is no clear observational evidence of separation between low- and high-luminosity LBVs. This is a result of our updated compilation of the fundamental parameters of LBVs, in which most LBVs have been analyzed with non-LTE radiative transfer atmospheric codes (mainly CMFGEN). A similar conclusion can also be drawn from the parameters quoted in \citet{clark05,clark09,clark12b}.

Obtaining the precise evolutionary status of observed LBVs is, however, a challenging task, because one has to rely on matching \lstar\ and \tstar\ of each individual star. Both quantities vary wildly in LBVs, on short timescales as a result of S-Dor type instability (changes in \rstar\ and \lstar\ without significant changes in \mdot\, see \citealt{ghd09}) and, more rarely, Giant Eruptions (ejections of several solar masses in a short outburst with increase in \lstar, see \citealt{hd94}). Both phenomena are not included in our models. In addition, LBVs may suffer from inflated radii \citep{grafener12a}. Taking the model predictions at face value, one would expect that LBVs with $\log \lstar < 5.3$ would be close to the end of their evolution, while those with $\log \lstar > 5.6$ would evolve to the WR phase. Given the uncertainties, LBVs with $5.3 < \log \lstar < 5.6$ could either come from $\mini>25~\msun$ and be at an intermediate stage of their evolution, or come from $\mini=20-25~\msun$ and be close to core collapse. It is worth noting that most LBVs have lower temperatures than the final stage of our 20-25~\msun\ models. This could point towards the observed LBVs having inflated radii, or having long S-Dor variability cycles, or not being at the endmost stage of their evolution. In addition, it could be also a result of the models not being tailored to match a particular star, as one could fine tune the rotation rate and mass lost at the RSG stage to bring the endpoints to lower temperatures.

In the above picture, one has to take into account the uncertain distance determinations for most LBVs, meaning that significant errors in $\lstar$ could exist. As an example, the LBV AG Car has had distance determinations in the range 2--6~kpc \citep{humphreys89}. A larger distance has been preferred, but it is mainly anchored on the kinematical determination based on the Galactic rotation curve \citep{stahl01,ghd09}. If AG Car were at a much closer distance of $\sim2-3~$kpc, its $\lstar$ would be much lower and fall within the range where AG Car would be an immediate SN progenitor. Stellar parameters of LBVs can nowadays be determined with reasonable accuracy with the advent of non-LTE radiative transfer codes such as CMFGEN. The speculative scenario above illustrates that constraining the fate of observed Galactic LBVs relies on more precise distance determinations, which may become available with the GAIA mission.

From a theoretical perspective, let us discuss the types of SN that would arise from stars that explode as LBVs according to our models, having in mind the aforementioned caveats associated with linking the chemical abundances of progenitors with SN types. In \citet{gme13}, we showed that the 20--25~\msun\ models have a very small amount of H ($\sim0.02~\msun$) left in their envelopes. This could point towards a SN IIb, meaning that LBVs could be the progenitors of SN IIb, such as SN 2008ax \citep{gme13}. However, until more firm constraints are put on the amount of H that characterizes the different types of SN II, we cannot rule out a type IIL. Interestingly, from an observational point of view, the original proposition that LBVs could be SN progenitors come from a study of transitional SNe and the effects of wind variability in their radio lightcurves \citep{kv06}. These were SN 2001ig (type II that evolved to Ib; \citealt{phillips01,clocchiatti02,fillipenko02}) and SN 2003bg (type Ic, later evolving to type II; \citealt{fillipenko03,hamuy03}). Therefore, there is observational support to our theoretical prediction that LBVs with relatively low luminosity are the progenitors of some SN IIb and other intermediate SN types between II and Ib.

Let us hypothesize what would happen in the case that a Giant Outburst occurred in the few years before the SN explosion of the  20--25~\msun\ models . This phenomenon is not included in our stellar evolution models, since the physical mechanism is unknown. As we discussed in \citet{gme13}, the progenitors  of 20-25~\msun\ rotating models, being LBVs and crossing the ``Yellow Void" in the HR diagram, could experience episodic mass loss. This would give rise to a dense circumstellar medium (CSM) characterized by relatively low velocities. The interaction of the SN blast wave with the CSM could then create a SN with narrow lines. The SN type would in principle depend on the amount of H in the ejecta, which in the case of our models is 0.02 and 0.00~\msun\ for the 20 and 25~\msun\ models, respectively. In the case of the rotating 20~\msun\ model, a Giant Outburst in the few years before the SN could produce a CSM with a range of relative H abundance values, depending on the amount of mass ejected. A SN IIn would arise if the amount of mass ejected in the outburst is small (up to a few tenths of a solar mass). In this case, H would still be sufficiently abundant in the ejecta to produce the typical H narrow lines seen in SNe IIn. For a more violent Giant Outburst, one would think that the fractional abundance of H in the ejecta would be very small, resulting later in a He-dominated spectrum and a SN Ibn. In the case of the 25~\msun\ rotating model, there is no noticeable amount of H in its ejecta. As a consequence, any Giant Outburst that happens at the pre-SN stage of this model would produce a He-rich CSM and an ensuing type Ibn SN, similar to SN 2006jc \citep{pastorello07}. 

Regarding the fate of the most massive stars ($\mini > 25~\msun$), we reinforce that our models of single stars predict that they would evolve to the WR phase. Since this is at odds with the observational suggestion that LBVs with  $\mini \simeq 50~\msun$ are the progenitors of type IIn SNe, let us examine the observational evidence of very massive LBVs exploding as SNe. Most of these come from the inference of large amounts of circumstellar material in the close vicinity of the progenitor star, such as in SN 2006gy \citep{smith07}. 

In a few cases, a very luminous progenitor ($M_V\sim -10$ mag) has been detected in pre-explosion images, such as the progenitors of SN 2005gl \citep{galyam09}, SN 2009ip \citep{smith10_sn2009ip,mauerhan12,foley11,pastorello12}, and SN 2010jl \citep{smith11_sn2010jl}. The high luminosity of the progenitor has been used as evidence for the star having initial masses  $\mini > 50~\msun$. In the case of the progenitors of SN 2009ip, the $M_V\sim -10$~mag measurement seems to correspond to a phase when the star was quiescent \citep{smith10_sn2009ip}. Thus, in this case, there is little doubt that the progenitor has $\mini > 50~\msun$, although one may wonder if the 2012 events observed in SN 2009ip correspond indeed to core collapse \citep{fraser13}. For the progenitors of SN 2005gl, however, there is no information about its pre-explosion variability. One may wonder if the measured magnitude in the pre-explosion image \citep{galyam09} was obtained during a Giant Outburst phase. During these Giant Outbursts, $M_V$ increases by 3 mag or more \citep{hd94} and, in this case, the progenitor of SN 2005gl could have had a much lower $M_V\sim -7$~mag during quiescence. This would imply an initial mass around 20-25~\msun  (Fig. \ref{figabsmagbc1}), which is consistent with our theoretical models of stars that explode during the LBV phase. 

We see thus that observational evidences for luminous LBVs as SN progenitors is mounting, but one may wonder whether this is the most usual evolutionary path of massive stars or a peculiar case of massive star evolution, perhaps resulting from binary evolution \citep{vanbeveren13}. There is no doubt that some progenitors of core-collapse SNe have their structure resulting from interactions in close binaries and these cannot of course be described by our single-star models. If indeed high luminous LBV star explode as SNe, then this type of SNe would be difficult to explain in the frame of the single star scenario, as can be seen in Fig. \ref{figabsmagbc1}. To obtain a clearer picture, a larger sample is needed, which may become possible with the current and next generation of transient surveys.

\section{Impacts on massive star evolution and concluding remarks}
\label{conc}
We showed in this paper that producing detailed spectra out of evolutionary tracks significantly impacts the current paradigm of stellar evolution and how different spectral types are linked to the final evolutionary stages. Here we focused on the pre-SN stage of single stars. Based on a stellar evolution perspective, we will present in a forthcoming paper the results of the spectral classification of the evolutionary model outputs throughout their full lifetime. Nevertheless, we are already able to assess important aspects of massive star evolution and fate, as we summarize below.

1. The current rotating models from our group \citep{ekstrom12} reproduce reasonably well the rates of different types of core-collapse SNe observed by \citetalias{smith11sn} and \citetalias{eldridge13}, with the exception of the SN IIP rate, which is overestimated. The rates of SN Ib and Ic predicted by the rotating models seem to be in line with the observations, although we note that the model rates may be overestimated, since we assume that all massive stars end their lives in a core-collapse SN event.

2. We performed combined stellar evolution and atmospheric modeling of massive stars at their pre-SN stage. With this approach, we are able to compute the emerging spectrum in high spectral resolution and perform synthetic photometry to obtain absolute magnitudes and bolometric corrections of the progenitors that have $\teff > 8000$ K. For the remaining progenitors, we supplement our analysis by using public MARCS models, scaled to the luminosity and interpolated in $\teff$ according to the end position in the HR diagram.

3. We found that massive stars, depending on their initial mass and rotation, end their lives as RSG, YHG, LBV, WN, or WO stars. For rotating models, we obtained the following types of SN progenitors: WO1--3  ($\mini \geq 32~\msun$), WN10--11 ($25 < \mini < 32~\msun$), LBV  ($20 \leq \mini \leq 25~\msun$), G1 Ia$^+$ ($18 < \mini < 20~\msun$), and  RSGs ($9 \leq \mini \leq 18~\msun$). For non-rotating models, we found the following spectral types of the SN progenitors: WO1--3  ($\mini > 40~\msun$), WN7--8 ($25 < \mini \leq 40~\msun$), WN11h/LBV  ($20 < \mini \leq 25~\msun$), and RSGs ($9 \leq \mini \leq 20~\msun$).

4. The most massive stars (initial mass above 32~\msun) end their lives with extremely high $\teff$ (150000--175000 K). Their spectrum is characterized by broad emission lines of \ion{C}{IV} and  \ion{O}{VI}, which are characteristic of WO stars. This is contrary to what has been widely expected based on the chemical abundance of the progenitor, which predicted a WC star at the pre-SN stage. Here we show that the high $\teff$, coupled with the presence of moderate amounts of O at the surface ($\sim30\%$ by mass), produces the morphology characteristic of WO 1--3 stars. For some WO models, we obtained spectral lines which are much weaker than observed in Galactic WOs, which may imply a hidden population of weak-lined WO stars or that our assumed values of $\mdot$ are too low. An increase in $\mdot$ by a factor of two is enough for producing a WO with an emission line strength in line with the observations.

5. We computed absolute magnitudes and bolometric corrections of core-collapse SN progenitors with initial mass in the range 9 to 120 ~\msun. The behavior of the absolute magnitudes and bolometric corrections as a function of initial mass is  regulated by how much flux from the star falls within the passband of a given filter. This depends on $\teff$, $\lstar$, $\mdot$, and $\vinf$. Therefore, not necessarily the most massive and luminous stars are the brighter ones in a given filter. We find that the RSGs are bright in the $RIJHK_S$ filters and faint in the $F170W$ and $UB$ filters. LBVs, YHGs, and WNLs are relatively bright and WOs are faint in all optical/IR filters, with the exception of the $F170W$ filter.

6. We obtained a relationship between the absolute magnitude of RSGs in different filters and their initial mass. This can be used to estimate the initial mass of progenitors detected in the pre-explosion images of SN IIP. Our method provides similar values of the initial mass of RSG as those found in the literature, given the uncertainties.

7. We discussed the detectability of SN Ib and Ic progenitors and argue that the WR stars that characterize the pre-SN phase are undetectable in the available pre-explosion images with the current magnitude limits. This is consistent with the current non-detection of WRs as progenitors of SNe in the available pre-explosion images of \citet{eldridge13}.

8. Unlike the SN IIP progenitors, we find that the absolute magnitude in a given filter of SN Ic progenitors do not depend strongly on the initial mass. As a consequence, in the event that a  type Ic progenitor is detected, our models indicate that it will be challenging to constrain its initial mass based on photometry (or spectroscopy) of the progenitor.

9. We predict that the limiting magnitudes in the observations should be at least 2 mag fainter than the current best limit in the $B$ band to detect single star progenitors of SN Ic at distances of 10--20~Mpc. For a SN 2002ap--like detection limit ($m_B=26.0$~mag, \citealt{crockett07}), the maximum distance that a SN Ic progenitor would be detected is $\sim 5.5$~Mpc. For typical detection limits ($m=24.5$~mag) and low amounts of extinction (0.3 mag), our models suggest that SN Ic progenitors from single stars would be detected up to a maximum distance  of $\sim 2.7$~Mpc.

\begin{acknowledgements}

We thank our referee, Jean-Claude Bouret, for a detailed reading and comments on the manuscript. We also thank John Hillier for making CMFGEN available, continuous support, and sharing models, and Jes\'us Ma\'{i}z-Appel\'aniz for making CHORIZOS available. We appreciate discussions with and comments on the manuscript from Simon Clark, Ben Davies, Luc Dessart, John Eldridge, John Hillier, Raphael Hirschi, Andreas Sander, and Jorick Vink. JHG is supported by an Ambizione Fellowship of the Swiss National Science Foundation. CG acknowledges support from EU- FP7-ERC-2012-St Grant 306901.

\end{acknowledgements}

\bibliographystyle{aa}
\bibliography{../../refs}

\end{document}